\documentclass[aps,pra,reprint,twocolumn,showpacs,showkeys,superscriptaddress,longbibliography]{revtex4-1}
\usepackage{amsmath,amssymb,amstext}
\usepackage[usenames,dvipsnames]{color}
\usepackage{graphicx}
\usepackage[english]{babel}
\usepackage{wasysym}
\usepackage{bm,bbold,braket}
\usepackage{natbib}
\usepackage{txfonts, comment, stmaryrd}
\usepackage{dcolumn}
\usepackage{graphicx}
\usepackage{subfigure}
\usepackage{float}
\usepackage[dvipsnames]{xcolor}
\usepackage[colorlinks=true, allcolors=blue]{hyperref}
\usepackage[normalem]{ulem} 

\newcommand{\ie}{i.\,e.,\ }

\newcommand{\fref}[1]{\text{Fig.}~\ref{#1}}
\newcommand{\ffref}[1]{\text{Figs.}~\ref{#1}}
\newcommand{\eref}[1]{\text{Eq.}~\eqref{#1}}
\newcommand{\eeref}[1]{\text{Eqs.}~\eqref{#1}}
\newcommand{\cms}[1]{{\color{black} #1}}

\begin{document}
	\title{Crystalline Phases of Laser-Driven Dipolar Bose-Einstein Condensates}
	
	\author{Chinmayee Mishra}
	\affiliation{Indian Institute of Technology Gandhinagar, Gandhinagar 382 355, India}   
	\author{Stefan Ostermann}
	\affiliation{Department of Physics, Harvard University, Cambridge, Massachusetts 02138, USA} 
	\author{Farokh Mivehvar}
	\affiliation{Institut für Theoretische Physik, Universität Innsbruck, Technikerstraße 21a, A-6020 Innsbruck, Austria}
	\author{B. Prasanna Venkatesh}
	\affiliation{Indian Institute of Technology Gandhinagar, Gandhinagar 382 355, India}

	\begin{abstract}
		Although crystallization is a ubiquitous phenomenon in nature, crystal formation and melting still remain fascinating processes with several open questions yet to be addressed. In this work, we study the \emph{emergent} crystallization of a laser-driven dipolar Bose-Einstein condensate due to the interplay between long-range magnetic and effectively infinite-range light-induced interactions. The competition between these two interactions results in a collective excitation spectrum with two roton minima that introduce two different length scales at which crystalline order can emerge. In addition to the formation of regular crystals with simple periodic patterns due to the softening of one of the rotons, we find that both rotons can also soften simultaneously, resulting in the formation of exotic, complex periodic or aperiodic density patterns. We also demonstrate dynamic state-preparation schemes for achieving all the found crystalline ground states for experimentally relevant and feasible parameter regimes.
	\end{abstract}
	
	\date{\today}
	\maketitle

     \section{Introduction}
    Ultracold atomic gases with long-range interactions are a platform with unprecedented properties to realize exotic many-body phenomena in a well-controlled environment~\cite{defenu_long-range_2021}. Long-range interactions in Bose-Einstein condensates (BECs) can either originate from the intrinsic magnetic dipole moment of atomic species~\cite{Lewenstein_2000,lahaye_physics_2009,Baranov2012,norcia_developments_2021,chomaz_dipolar_2022}, or be imposed by manipulating the BEC with external laser fields~\cite{Kurizki_2002, Kurziki_2003,honer_collective_2010,Ritsch_2016,ostermann_probing_2017,dimitrova_observation_2017,Zhang2018,Zhang2021,Chatterjee2018,Chatterjee2022} or quantized dynamic cavity fields~\cite{Esslinger_2012,ritsch_cold_2013, Mivehvar2021Cavity,Karpov2022}.
	In addition to supersolid and crystalline (droplet array) phases precipitated by the long-range interactions, competition between diverse interactions in such systems can also lead to other interesting emergent physics. These include frustration in BECs confined to multi-mode cavities \cite{Lev1,Lev2,Lev3} and quasi-crystalline order in dipolar BECs with spin-orbit interactions \cite{Deng2012,Sarang2013,Li_2019} or non-dipolar BECs interacting with multiple cavities \cite{Mivehvar2019}.
	
	\begin{figure}[b!]
		\centering
		\includegraphics[width = \columnwidth]{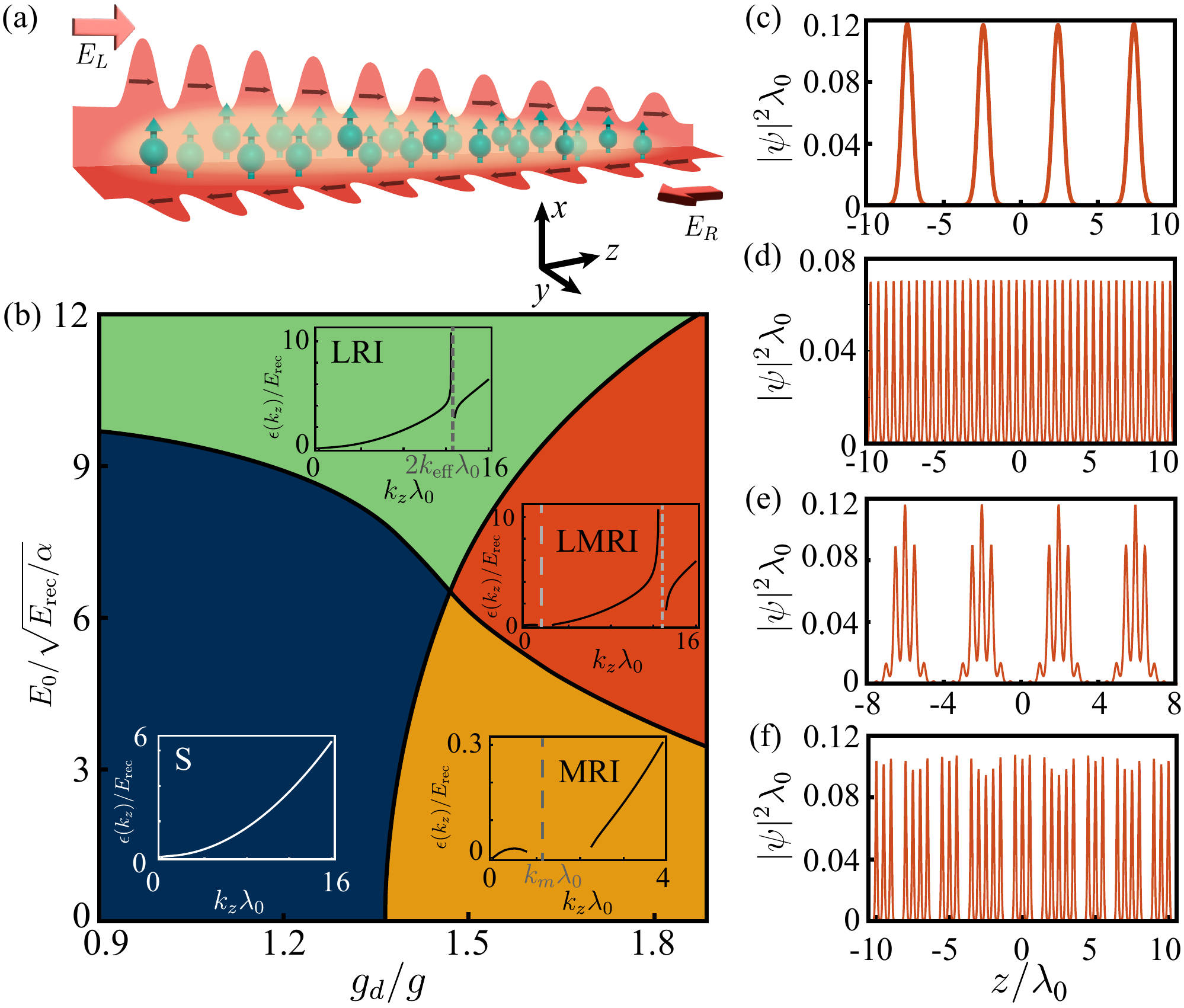}
		\caption{(Color online) (a) Schematic of a dipolar BEC in the presence of two counter-propagating laser beams of orthogonal polarisation. (b) Stability diagram of a \emph{homogeneous} BEC as a function of dipole-interaction strength $g_d$ and light amplitude $E_0$ calculated from the excitation spectrum. The insets in the four regimes show examples of the typical spectrum that is stable (S) or consisting of either light (LRI) or magnetic (MRI) roton instability, or \emph{bi}-roton instability (LMRI). \cms{The gray dashed lines are a guide to the eye for the location of any roton instabilities.} (c-f) Atomic density patterns in the crystalline ground state phases. (c) Magnetic droplet crystal (MC) for $\{g_d/g, \sqrt{\alpha} E_0/\sqrt{E_{\text{rec}}}\}=\{1.52, 0\}$ and (d) light crystal (LC) for $\{g_d/g,\sqrt{\alpha}E_0/\sqrt{E_\text{rec}}\}=\{0.92,12\}$. (e-f) Light-magnetic crystal (LMC) for $\{g_d/g, \sqrt{\alpha} E_0/\sqrt{E_{\text{rec}}}\}=\{1.42, 4.2\}$ and $\{1.7, 5\}$, respectively. Panel (e) corresponds to `droplets of supersolids' and (f) to an aperiodic `crystal'. Common parameters for (b-f) include $L=50\lambda_0$, $\omega_\rho = 100 E_{\rm rec}/\hbar$, $\zeta=0.1$, $a=70 a_0$ ($a_0$-Bohr radius) and $N=10^5$ atoms.}
		\label{fig1}
	\end{figure}

	In this paper, we focus on the interplay between long-ranged magnetic dipole and effectively infinite-ranged light-induced interactions in a cigar-shaped elongated BEC illuminated by two counterpropagating laser beams with orthogonal polarizations [see~\fref{fig1}(a)]. In the absence of light this system is expected to exhibit a phase transition to supersolid~\cite{Ferlaino_2019, bottcher_transient_2019, Pfau_2019, Modugno_2019} or droplet crystalline phases~\cite{Santos_2016, Blakie_2016}. Alternately, it has been shown that for a non-dipolar, laser-driven BEC the translation invariance of the system can be broken, leading to the simultaneous formation of a crystalline atomic state and optical potential with an intrinsically chosen period comparable to that of the laser field's wavelength~\cite{Ritsch_2016, ostermann_probing_2017, dimitrova_observation_2017}. The formation of these phases is related to the instability of a magnetic~\cite{Ferlaino_2018, Natale_2019} or a light-induced roton mode in the excitation spectrum~\cite{Ritsch_2016, ostermann_probing_2017, dimitrova_observation_2017} respectively, similar to the one originally predicted for superfluid helium-4~\cite{landau_theory_1941}. 
	
	The fundamental question we pose here is: what are the phases that emerge from the competition between these two distinct interactions? We show that this comprises an intriguing scenario leading to the formation of a rich variety of crystalline and supersolid phases. Specifically, from the collective excitation spectrum we find regimes with \emph{bi-}roton softening arising from the competition between the two long-range interactions, indicating the existence of two possible crystallization length scales. We confirm this by calculating the ground-state phase diagram which hosts, besides the two individual ordered states corresponding to each long-range interaction, an intertwined emergent phase with periodic or aperiodic density patterns corresponding to the \emph{bi}-roton softening. Ultimately, we outline state preparation schemes to achieve the different crystalline ground-state phases dynamically for experimentally feasible conditions.

   \cms{The paper is organized as follows. In section \ref{sec:model} we describe the system and set up the governing equations. In section \ref{sec:colexc} we analyze the elementary excitations of a uniform condensate to demonstrate the emergence of the bi-roton spectrum. Section \ref{sec:percry} highlights the unique density modulations characterizing the crystalline ground states we obtain as a direct consequence of the various instabilities arising in the spectrum. In section \ref{sec:phase} we delineate the phase diagrams in terms of different observables demarcating the domains associated with various crystalline states found. Finally, the state preparation dynamics have been detailed in section \ref{sec:dyn}. We provide some additional details that supplement the discussion in the paper in appendices \ref{app:hhe}-\ref{app:heating}.}
	
	\section{Model}\label{sec:model}
	We consider a dipolar BEC at zero temperature confined by a transverse harmonic trap with frequency $\omega_\rho$ into a cigar shaped geometry along the $z$-direction [see~\fref{fig1}(a)]. The magnetic dipoles are oriented along the $x$-direction. In addition, the BEC is subject to two counter-propagating, far-off resonant and orthogonally polarized (i.e., non-interfering) plane-wave laser beams. For atoms (with mass $m$) confined by an axial box potential $V_{\text{box}}(z)$ of extent $L$, the BEC order parameter is decomposed as $\Psi(\textbf{r},t)=\psi(z,t)e^{-(\eta x^2 + y^2/\eta)/2l^2}/(\sqrt{\pi}l)$ where the transverse width(anisotropy) $l$($\eta$) remains a variational parameter following the reduced 3D theory~\cite{Blakie_2020, Ferlaino_2020}. The dynamics of $\psi(z,t)$ is governed by the extended Gross-Pitaevskii equation including the Lee-Huang-Yang (LHY) correction term~\cite{Santos_2016, Blakie_2016},
	\begin{align}
		i\hbar \dot{\psi} & = \left [\mathcal{E}_\rho-\dfrac{\hbar^2\nabla_z^2}{2m} +V(z)+  \Phi_\rho(z) + g_{\text{LHY}}N^{3/2}|\psi|^3 \right] \psi
		\label{eq:eGPE}
	\end{align}
	with $\int dz|\psi(z,t)|^2 = 1$. The interaction term is given by 
    \begin{align}
        \Phi_\rho = \dfrac{gN}{2\pi l^2} |\psi|^2+\dfrac{g_dN}{2\pi l^2}\int dk_z e^{i z k_z} V_{d}(k_z) n(k_z)
    \end{align}
    with the first term representing the short-range interaction of strength $g=4\pi\hbar^2 a /m$ ($a$ denoting the s-wave scattering length) and the second term the dipole-dipole interaction (DDI) with magnitude $g_d = \mu_0 d^2/3$ for atoms with a dipole moment $d$. Furthermore, $n(k_z)$ is the Fourier transform of the density and $V_d(k_z)$ is the dipole interaction in momentum space given by $V_d(k_z) = [3(1-q^2e^{q^2}\Gamma[0,q^2])/(1+\eta)-1]$ with $q=k_z^2l^2\sqrt{\eta}/2$ and $\Gamma[a,b]$ denoting the incomplete Gamma function. The transverse energy $\mathcal{E}_\rho = (\hbar^2/4ml^2+ml^2\omega_\rho^2/4)(\eta+1/\eta)$. The magnitude of the LHY correction term $\propto |\psi|^3$ is given by $g_{\text{LHY}}=(64g a^{3/2}/15 \pi^2 l^3)(1+3g_d^2/2g^2)$.
	
	The potential $V(z)$ in~\eref{eq:eGPE} consists of $V_\mathrm{box}(z)=0 ~\mathrm{if} ~\vert z\vert \leq L/2, ~\mathrm{else}~ \infty$, and $V_\mathrm{opt}(z)$ induced by the incoming light beams,~\ie $V(z)=V_{\mathrm{box}}(z)+V_{\mathrm{opt}}(z)$. The optical potential depends only on the sum of the individual intensity distributions of the left $E_L(z)$ and right $E_R(z)$ propagating laser fields as $V_{\text{opt}}(z)=-\alpha\left(|E_L(z)|^2+|E_R(z)|^2\right)$, with $\alpha$ denoting the real part of the polarizability of the atoms. The laser fields individually satisfy the Helmholtz equation with the atomic density acting as a refractive medium,
	\begin{eqnarray}
		\dfrac{\partial^2}{\partial z^2}E_{L,R}(z) + \dfrac{(2\pi)^2}{\lambda_0^2}\left[1+\zeta\lambda_0|\psi(z,t)|^2\right]E_{L,R}(z)=0,
		\label{hhe}
	\end{eqnarray}
	subject to appropriate boundary conditions (see  Appendix \ref{app:hhe}). Here, $\lambda_0 = 2 \pi/k_0$ denotes the wavelength of the incoming plane-wave laser field. The dimensionless quantity $\zeta=\alpha N/2\pi\epsilon_0 l^2\lambda_0$ characterizes the coupling between the atomic density and the light. Note that for running-wave laser fields in the absence of the atomic back-action, $V_{\text{opt}}(z)$ amounts simply to a position-independent constant energy shift.

    The coupled \eref{eq:eGPE} and \eref{hhe} have to be solved in conjunction with the minimization of the following energy functional with respect to the parameters $l,\eta$ that determine the full $3$D order parameter $\Psi(\textbf{r},t)=\psi(z,t)e^{-(\eta x^2 + y^2/\eta)/2l^2}/\sqrt{\pi l}$:
    \begin{align}
        \mathcal{E}(\psi;l,\eta) &= \mathcal{E}_\rho + \int dz\,\psi^*(z,t) \left[-\dfrac{\hbar^2}{2m}\nabla^2+V(z)\right.\nonumber\\
        & \left.\hspace{2cm} +\frac{\Phi_\rho}{2}+\dfrac{2g_{\text{LHY}}N^{3/2}}{5}\vert \psi \vert^3\right]\psi(z,t) \label{enfunc}.
    \end{align}
    The method used to solve the Helmholtz equations~\cite{Ritsch_2016} is covered in Appendix \ref{app:hhe}.
	
	\section{Collective excitations and instabilities of a homogeneous condensate.}\label{sec:colexc}
     To understand the nature of the ground states of the coupled Eqs.~\eqref{eq:eGPE} and \eqref{hhe} in the absence of $V_\text{box}$, we analyze the collective excitation spectrum of the system by linearizing the equations of motion about a homogeneous atomic wavefunction $\psi_0(z) = 1/\sqrt{L}$ and plane-wave fields $E^0_{L,R}(z)=E_0 e^{\pm i k_{\mathrm{eff}}z}$ with $E_0$ denoting the amplitude of the driving laser fields far away from the BEC. The effective propagation number $k_\mathrm{eff}=2\pi \sqrt{1+\zeta \lambda_0|\psi(z)|^2}/\lambda_0$ in a homogeneous atomic cloud. We can write:
    \begin{eqnarray}
    	\psi(z) &=& \left[\psi_0(z) +u e^{-i(k_z z-\omega t)}+v^*e^{i(k_z z-\omega t)}\right]e^{-i\mu t/\hbar},\nonumber\\
    	E_{L,R}(z)&=& E^{0}_{L,R}(z) + \delta E.\nonumber
    \end{eqnarray}
    Using the above ansatzes in \eref{eq:eGPE} and \eref{hhe} and keeping terms up to linear order in the fluctuations $\delta E, u,v^{*}$, the calculations are easily performed in Fourier space. The expression for $\delta E$ after reverting back to position space is given by,
    \begin{eqnarray}
	\delta E &=& - \dfrac{(2\pi)^2 \zeta E_0}{\lambda_0 \sqrt{L}}\left[\dfrac{(u+v)e^{-i(k+k_\mathrm{eff})+i\omega t}}{k_\mathrm{eff}^2-(k+k_\mathrm{eff})^2} +\dfrac{(u^*+v^*)e^{i(k-k_{\mathrm{eff}})-i\omega t}}{k_\mathrm{eff}^2-(k-k_\mathrm{eff})^2}\right]\nonumber.
    \end{eqnarray}
    Using the above expression, the eGPE is linearized in a standard way. The resulting dispersion relation reads,
	\begin{eqnarray} \label{eq:elementary-excitation}
		\epsilon(k_z) &= \left[\dfrac{\hbar^2 k_z^2}{2m} \left \{ \dfrac{\hbar^2 k_z^2}{2m}+\dfrac{gN}{\pi l^2 L}+\dfrac{g_dN}{\pi l^2 L}V_d(k_z)+\dfrac{3g_{\text{LHY}}N^{3/2}}{L^{3/2}}\right.\right.\nonumber\\
		&\left.\left. -\dfrac{32 \pi^2\zeta\alpha|E_0 |^2}{L\lambda_0(k_z^2-4k_\mathrm{eff}^2)}\right\}\right]^{1/2}\label{spec},
	\end{eqnarray}
     Clearly, the spectrum has features from both the magnetic DDI and light-induced interactions (LII). Note that the terms corresponding to the interactions and the quantum fluctuation have a dependence on the variational parameters $l$ and $\eta$, which are obtained from the minimization of a reduced form of the energy functional in \eref{enfunc},
    \begin{align}
	\mathcal{E}_\mathrm{hom}&=\mathcal{E}_\rho +\dfrac{g N}{4\pi l^2 L}+\dfrac{g_d N}{4\pi l^2 L}\left(\dfrac{3}{1+\eta}-1\right) \nonumber\\
 &\hspace{2cm}+ \dfrac{2}{5} \dfrac{g_\mathrm{LHY}N^{3/2}}{L^{3/2}} - 2 \alpha E_0^2.
    \end{align}

    \cms{By looking at the Eq.~\eqref{spec} it is clear that there exists a singularity at $k_z = 2 k_{\text{eff}}$ which sets the LC periodicity ~\cite{Ritsch_2016}. In principle, this divergence is compensated by the infinitely large $L$ where the reflection of incident light at the edge of the condensate boundary can be neglected which is a purely finite-sized effect incorporated in our model. The singularity is easily avoided by considering finite $L$ which enforces quantization of the momentum values i.e., $k_z$ can only take discrete values in the multiple of $2\pi /L$. The divergence in the truly infinite $L$ limit is a limitation of the model which can be overcome when the retardation effects of the light fields are taken into account.}
	
	In~\fref{fig1}(b), we show the three distinct types of instabilities [$\epsilon^2(k_z)<0$] which can occur in this system, according to Eq.~\eqref{eq:elementary-excitation}, as a function of the strength of magnetic dipolar interaction $g_d/g$ and the amplitude of the light fields $E_0$~\cite{Note2}. Insets in~\fref{fig1}(b) show the representative spectrum for each parameter region. We see that for a fixed small value of $\sqrt{\alpha}E_0/\sqrt{E_\text{rec}}\lesssim 6$ the system develops a magnetic roton as $g_d/g$ is increased. This roton eventually softens at the wavenumber $k_\mathrm{m}$, signaling a transition from the stable (S) regime to the magnetic roton instability (MRI) regime~\cite{Ferlaino_2018}. Similarly, increasing $E_0$ at a fixed and sufficiently small $g_d/g \lesssim 1.3$ leads to a roton induced by the light fields. This roton softens at $2k_{\text{eff}}$ to enter into the light roton instability (LRI) regime~\cite{Ritsch_2016}. 
	
	In addition to these expected instabilities where one of the two long-range interaction is dominant, we also find a third type of instability when both $g_d/g$ and $E_0$ are comparatively strong to enter a \emph{bi-}roton instability (LMRI) region. Here, the magnetic and the light-induced rotons are simultaneously unstable \cite{Note2a}. As we discuss later, contributions from both wavenumbers $k_\mathrm{m}$ as well as $2k_{\text{eff}}$ $(>k_\mathrm{m})$ give rise to a new phase with periodic or aperiodic density patterns. Moreover, the non-linearity of the phase boundaries in~\fref{fig1}(b) clearly shows the interplay between the rotons. The S-LRI and MRI-LMRI transition boundaries are significantly altered when $g_d/g$ is increased as the nonlinear dependence of DDI on $k_z$ helps soften the higher momentum modes and lower the critical $E_0$ needed to instigate the transition. Alternately, an increase in $E_0$ pushes the S-MRI and LRI-LMRI transition boundaries to higher $g_d/g$ values as the light fields counteract the unstable magnetic roton and cure it. This can be well understood by the low momentum behaviour of the spectrum $\epsilon(k_z<<2k_{\text{eff}})$ where the last term in~\eref{spec} becomes dominantly positive requiring higher magnitude of $g_d/g$ for magnetic roton softening. An important distinction between the two rotons is while LRI remains sharply peaked at $2k_\text{eff}$, the MRI can span over a broad range of momenta. This greatly influences the density distribution of the corresponding ground states.

	\section{Periodic and aperiodic crystallization}\label{sec:percry}
     In order to obtain the density-wave ground states precipitated from the various roton instabilities, we look for the stationary states of the system in the potential $V_\text{box}$ of finite extent $L$~\cite{Note3}. We employ imaginary time evolution and conjugate gradient methods~\cite{Tang_2017, Bohn_2006} along with a fourth-order Runge-Kutta method to simultaneously solve the eGPE and the Helmholtz equation, Eqs.~\eqref{eq:eGPE} and \eqref{hhe}. Deep in the MRI and the LRI regimes a straight-forward mapping exists to the stationary states of the magnetic crystal (MC) [see~\fref{fig1}(c)]~\cite{Ferlaino_2018} and light crystal (LC) [see~\fref{fig1}(d)]~\cite{Ritsch_2016} phases, respectively. The periodicity of these density patterns for the MC (LC) is set by the softened momenta associated with the magnetic (light) roton.
	
	Apart from these two known phases, the \emph{bi}-roton instability engenders peculiar light-magnetic crystal (LMC) states, where the two long-range interactions compete with one another. This can either result in density waves with periodic or aperiodic order [see~\fref{fig1}(e,f)]~\cite{Wolff_1982}.
	\fref{fig1}(e) shows an example of the former. The density exhibits a periodic envelope of droplets (induced by the DDI) where each of them support intra-droplet crystals (set by the LII) of smaller periodicity, thus forming a unique `droplets of supersolids' state. Note that the parameters $g_d/g=1.42,\sqrt{\alpha} E_0=4.2\sqrt{E_{\text{rec}}}$ used for~\fref{fig1}(e) indicate that the LII effects are prominent even below the LMRI threshold in~\fref{fig1}(b) due to finite-size effects as discussed further below. This is in contrast to the aperiodic ordered pattern in~\fref{fig1}(f) that is observed for higher $E_0$ values away from the MRI-LMRI boundary. The lack of discrete translational symmetry in such structures can be attributed to contributions from a broad range of momenta associated with the softened magnetic roton. Furthermore, note that the emergence of atomic density patterns shown in \ffref{fig1}(d-f) is accompanied by the development of a standing-wave light field~\cite{Ritsch_2016} \cms{(see Appendix A)}.
	
	\begin{figure}
		\centering
		\includegraphics[width = \columnwidth]{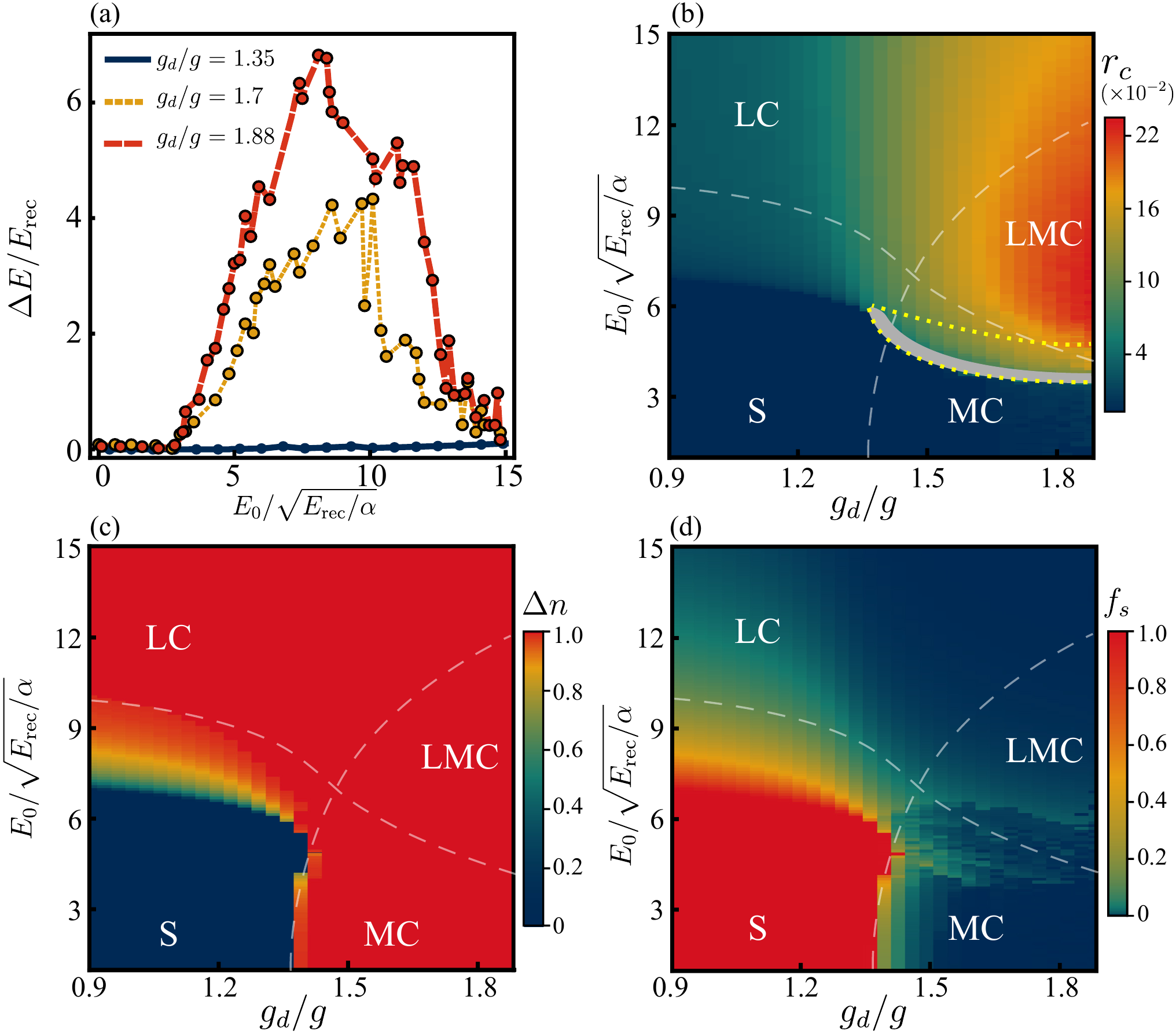}
		\caption{(a). Maximum energy difference between converged states of the eGPE for six different initial guesses for three $g_d/g$ values. (b-d). Mean-field phase diagram as a function of $\{g_d/g, \sqrt{\alpha}E_0/\sqrt{E_{\rm rec}}\}$ characterized by reflection coefficient $r_c$ (b), density contrast $\Delta n$ (c), and superfluid fraction $f_s$ (d). White dashed lines are the stability diagram boundaries from the excitation spectrum. Yellow dotted curve demarcates the `droplets of supersolid' phase and gray strip highlights the domain where density patterns are always periodic. All other parameters are the same as~\fref{fig1}(b).}
		\label{fig2-3}
	\end{figure}
	
	\begin{figure*}
		\centering
		\includegraphics[width = .95
		\textwidth]{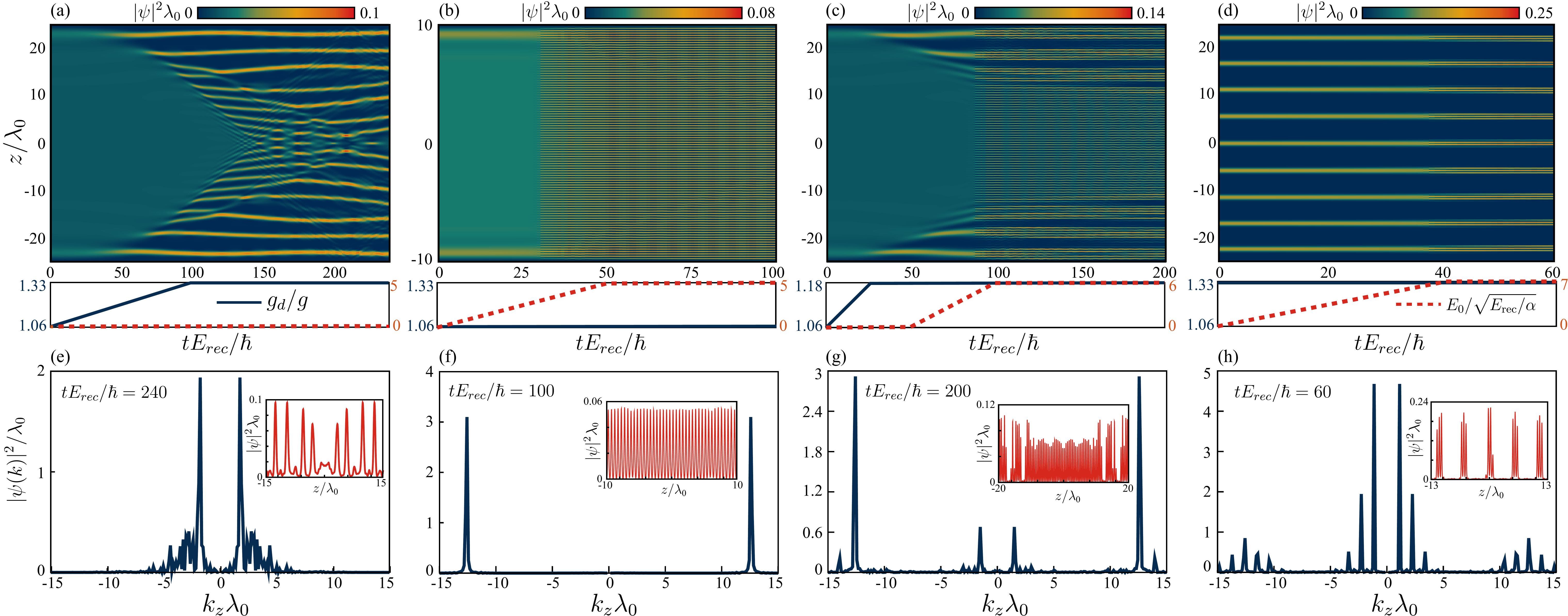}
		\caption{Preparation of (a) magnetic crystal (MC), (b) light crystal (LC), and (c) light-magnetic crystal (LMC) states starting from an unordered state, and (d) LMC starting from a MC for Dy atoms with their corresponding ramping schemes of $g_d/g$ and the light intensities $E_0$ shown in the middle row. The bottom row (e-h) depicts the corresponding densities in momentum space at the final time. For concreteness, the final-time density distributions in real space are shown in the insets. All other parameters are the same as~\fref{fig1}(b).}
		\label{fig4}
	\end{figure*}
	
	
	\section{Phase diagram} \label{sec:phase}
     The biggest challenge in determining the numerical phase diagram of this system is the highly non-convex nature of the energy landscape in the regimes with strong DDI, where the simulations converge to different local minima for different initial guesses irrespective of the numerical methods used. For purely magnetic crystals this is taken care of by starting from different multi-Gaussian ansatzes and comparing their final energies~\cite{Blakie_2018}. However, when the applied light fields $E_0$ are also increased, not only the energetically dense local minima are potentially numerous but also the choice of initial guesses is no longer obvious. Therefore, convergence to the true global minimum remains ambiguous. In~\fref{fig2-3}(a) we parameterize the non-convexity of the energy landscape via the maximum energy difference $\Delta E$ obtained from different initial ansatzes. Deviation of $\Delta E$ from zero indicates that the converged solutions are ``quasi-stationary'' states, associated with different local minima. The recovery of convexity in the energy landscape for high $E_0$ can be intuitively attributed to the `curing' of unstable magnetic roton due to increasing LII.
	
	Interestingly, the non-convexity does not hinder the detection of the phase boundaries as the qualitative nature of the density patterns obtained from all initial guesses remains same \cms{(we provide the expressions for the different initial guesses used in Appendix B)}. To obtain the structural transition boundaries we focus on three key observables -- namely, the reflection coefficient $r_c$ \cms{(defined in Appendix A)} which measures the back-reflection of the incoming light fields due to the dynamic formation of a density grating, the density contrast $\Delta n= |n_{\text{max}}-n_{\text{min}}|/(n_{\text{max}}+n_{\text{min}})$ in the bulk of the condensate, and the superfluid fraction $f_s = (L/n)(\int |\psi|^{-2} dz)^{-1}$~\cite{Leggett_1970, Rica_2008, Ferlaino_2020}. The mean-field phase diagrams are shown in~\fref{fig2-3} in the parameter space of $g_d/g$ and $E_0$.
	
	The reflection coefficient $r_c$ acts as a robust parameter to detect the onset of dominant light effects and as a non-destructive experimental probe of the emergent light crystalline (LC) and light-magnetic crystalline (LMC) order. Although the excitation spectrum boundaries are in qualitative agreement with the numerical simulations, the influence of the box-potential-induced edge effects lowers the threshold $E_0$ at which LII effects can become prominent as seen in~\fref{fig2-3}(b). It is further lowered at $g_d/g\gtrsim 1.35$ when strong DDI leads to increasingly denser droplets. This effect is captured by the increase in $r_c$ as the medium gets more opaque from LC to LMC. Additionally, the structural transition curve from MC to LMC is much steeper than that from S to LC due to higher $g_d/g$. We find that this transition region precisely hosts the `droplets of supersolid' phase and demarcates the same in~\fref{fig2-3}(b).  Furthermore, the non-convexity of the energy landscape can also be seen in the behavior of $r_c$ \cms{(see Appendix A)}. The contrast $\Delta n$ is used to faithfully mark any transition from the S to the crystalline phases (LC, MC, and LMC) and varies smoothly across the crystalline phases; see \fref{fig2-3}(c). 
	
	Lastly, the superfluid fraction $f_s$ shown in \fref{fig2-3}(d) reduces as $g_d/g$ increases for any fixed $E_0$ in LC and LMC phases. For $g_d/g\gtrapprox1.5$, there is a recovery of the superfluidity as one enters the LMC phase from MC accompanied by the emergence of the peculiar droplet of supersolid states. Clearly, light-induced interactions play a significant role in enhancing the supersolid properties. This is further validated by evaluating the phase coherence \cite{Bland2022} in numerical simulations of the state preparation including thermal noise \cms{(see Appendix C)}.
	
	\section{State preparation and dynamics}\label{sec:dyn}
     Finally, we demonstrate in~\fref{fig4} that despite the non-convexity of the energy landscape all crystalline phases can be prepared dynamically. This is in contrast to a recent work where the combination of non-convexity and symmetry leads to amorphous behaviour for a self-organized BEC in a cavity with Rydberg-excitation-induced long-range interactions \cite{Yelin_2021}. For the experimentally relevant $d=10\mu_B$ (Dy atoms) and $\omega_\rho/2\pi=100$ Hz, beginning with a uniform bulk condensate at $a=100a_0$ and $E_0=0$ in the S phase, the three different crystal phases are obtained by either quenching $g_d/g$ (by varying $a$) or ramping $E_0$ up in a box trap. The sweeping schemes are plotted in the second row of \fref{fig4}(a-c). 
	
	The dynamics reveal several crucial distinctions between the emergence of the MC [\fref{fig4}(a)] vs the LC [\fref{fig4}(b)]. In MC, the crystalline order sets in locally from the edges \cite{Recati_2022} and grows towards the center while in LC, the onset of order is sharp and global. The MC excitation due to the sweeping involves both lattice vibrations and amplitude oscillations. In contrast, the phononic modes are almost frozen for the LC once it sets in. This pinning effect is a by-product of the singularly dominant momentum peak at $\pm2k_{\text{eff}}$ as well as the light-field boundary conditions. 
	
	During the emergence of the LMC phase [see~\fref{fig4}(c)] both these behaviors are observed as $g_d/g$ and $E_0$ are swept sequentially. Interestingly, depending on the holding time after $g_d/g$ is quenched and before $E_0$ is ramped, very different LMC density patterns can be obtained due to the pinning effect. In~\fref{fig4}(d) the MC is chosen as the initial state, as opposed to the homogeneous state in~\ffref{fig4}(a,b,c). This provides a greater control over the desired LMC state. For example, the droplets of supersolid state can be prepared by ramping $E_0$ which allows the deterministic manipulation of the intra-droplet contrast. The momentum space distribution and spatial patterns (inset) of the densities at final times are illustrated in the bottom row of~\fref{fig4}. A visual comparison between~\fref{fig4}(e) and (f) clearly shows that the fat-tailed distribution in case of the MC correlates with the `softness' of the crystalline order while the single peak corresponding to the LC indicates the `stiffness' of the spatially pinned LC~\cite{Guo_2021}. In \cms{Appendix C} we have supplemented the ideal state preparation dynamics presented here with those including thermal noise and found good qualitative agreement between them.
	
	\section{Conclusions and outlook}
     In conclusion, we have demonstrated that competing long-range interactions in a laser-driven dipolar BEC can lead to a rich phase diagram with a variety of crystalline phases. An important challenge to realize the predicted crystalline structures is to minimize the laser-induced heating rate which scales as $R \sim ({\Gamma^3}/{8\Delta_a^2})({I}/{I_{\text{sat}}})$ for an optical transition with linewidth $\Gamma$, detuning $\Delta_a$, saturation intensity $I_{\text{sat}}$ and laser intensity $I = c \epsilon_0 \vert E_0\vert^2/2$. We show in detail in \cms{Appendix E} that this heating rate takes manageable values in state-of-the-art setups with Erbium or Dysprosium (Dy) BECs. For instance, the $741$~nm transition of Dy with $\Gamma = 2 \pi \times 1.8$~KHz, with a laser intensity $I = 0.6$~W/cm$^2$, and detuning $\Delta_a = 2 \pi \times 1.6$~MHz leads to $R \sim 34$~Hz. Comparing this to our state preparation time scales of $\sim 100\hbar/E_\mathrm{rec}$, it becomes clear that the phases we predict are achievable in current experimental setups. Our work also opens up a promising direction for next-generation experiments and theoretical studies involving dipolar BECs where the addition of a laser drive leads to fascinating phenomena. Some pertinent follow-up questions, to be addressed elsewhere~\cite{FollowupPRA}, include a detailed analysis of the phase coherence in the LMC phase (beyond what is presented in \cms{Appendix C}) and the impact of harmonic trapping along the axial $z$ direction.
	
	\begin{acknowledgments}
		C.~M.\ acknowledges support from IIT Gandhinagar via the Early Career Fellowship program. S.~O. is supported by a postdoctoral fellowship of the Max Planck Harvard Research Center for Quantum Optics. F.~M.\ acknowledges financial supports from the Stand-alone project P~35891-N of the Austrian Science Fund (FWF), and the FET Network Cryst3 funded by the European Union (EU) via Horizon 2020. F.~M.\ and B.~P.~V.\ acknowledge support from an India-Austria DST-BMWF joint project with the project numbers IN~05/2020 and DST/INT/BMWF/AUSTRIA/P-06/2020, respectively. 
	\end{acknowledgments}
\appendix


\begin{figure*}
	\centering
	\includegraphics[width = \textwidth]{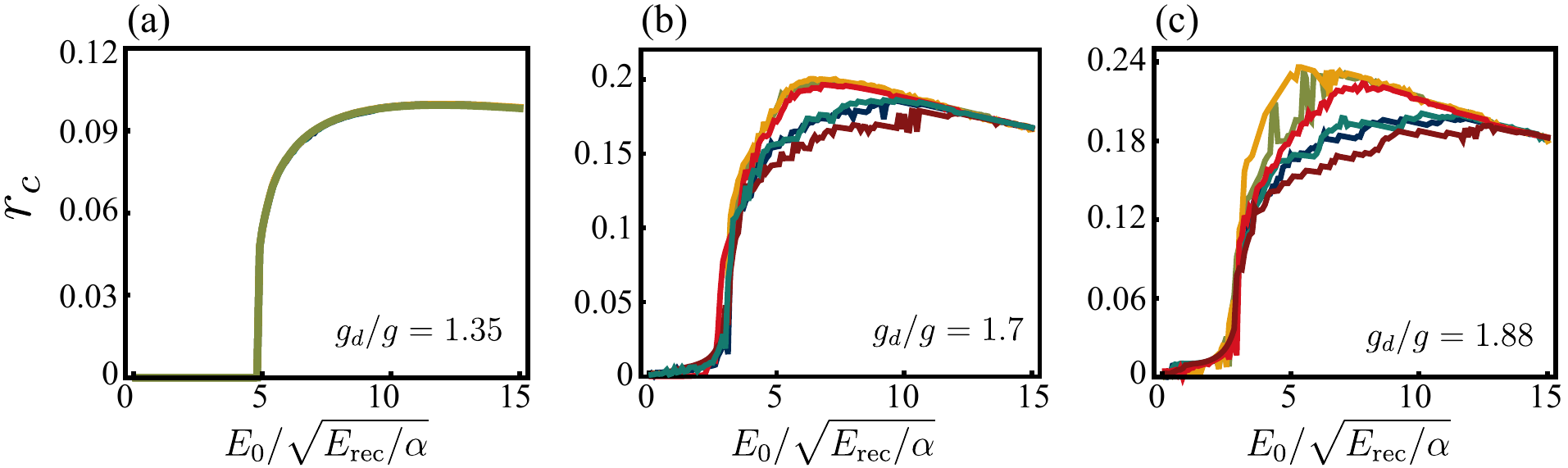}
	\caption{Reflection coefficients $r_c$ for converged states of eGPE obtained from different initial guesses for $g_d/g = 1.35$ (a), $1.7$ (b) and $1.88$ (c) corresponding to the same parameters used in Fig.~1(b) of the main text.}
	\label{sfig2}
\end{figure*}

\section{Helmholtz Equation Solution}\label{app:hhe}
We now detail the procedure to solve the Helmholtz equation specified by Eq.~(2) of the main paper (or in Eq.~(7) of the main paper appendix in a dimensionless form) for a given condensate order parameter $\psi(z,t)$ inside the finite sized box potential extending from $-L/2<z<L/2$. Consider the incident beam on the BEC of size $L$ from left. The boundary conditions to solve the Helmholtz equation for either the left or the right propagating light field within the BEC can be determined by first recognizing that the light field to the left of the condensate is given by $E^{\text{left}}(x) = A e^{i k_0 (x+L/2)} + B e^{-i k_0 (x+L/2)}$ and field to the right denoted by $E^{\text{right}}(x) = D e^{i k (x-L/2)}$ \cite{Ritsch_2016}. The relation between the incident ($A$), reflected ($B$), and transmitted ($D$) amplitudes is given by:
\begin{eqnarray}
	B &=& r_c A \label{eq:rcDefn}\\
	D &=& t_c A \label{eq:tcDefn}
\end{eqnarray}
and defines the reflection and transmission coefficients $r_c$ and $t_c$ respectively. Note that $\vert r_c \vert^2 + \vert t_c \vert^2 = 1$. The electric field at the boundary of the BEC is given by:
\begin{align}
	E^{\text{left}}(-L/2) = A + B, & \hspace{0.3in}\frac{\partial E^{\text{left}}}{\partial z}(-L/2) = i k_0 (A - B), \label{eq:CauchyBCs} \\
	E^{\text{right}}(L/2) = D , & \hspace{0.3in} \frac{\partial E^{\text{right}}}{\partial z}(L/2) = i k_0 D. \nonumber
\end{align}
In order to compute $r_c$ we take an arbitrary value for the incident amplitudes $E^{\text{left}}(-L/2)$ and $\frac{\partial E^{\text{left}}}{\partial z}(-L/2)$, as the Cauchy boundary condition and solve the Helmholtz equation in the region $-L/2\leq z \leq L/2$ using the fourth order Runge-Kutta method. This allows us to determine $E^{\text{right}}(L/2)$ and $\frac{\partial E^{\text{right}}}{\partial z}(L/2)$. From the ratios $r_1 = E^{\text{left}}(-L/2)/E^{\text{right}}(L/2)$ and $r_2 = \dot{E}^{\text{left}}(-L/2)/\dot{E}^{\text{right}}(L/2)$ (where $\dot{E} = \frac{\partial E}{\partial z}$), one obtains the reflection coefficient as
\begin{eqnarray}
	r_c = \dfrac{r_1 - r_2}{r_1 + r_2},
\end{eqnarray}
for a given atomic order parameter $\psi(z,t)$. Once we have $r_c$, we can now set the amplitude of the incident light as the laser driving field amplitude $A = E_0$ and solve the Helmholtz equation with boundary conditions given by \eeref{eq:rcDefn}-\eqref{eq:CauchyBCs} to determine $E_L(z)$. A similar approach can be used to solve for $E_R(z)$ using the light beam incident from the right. Since we have only considered symmetric driving strength from the left and right, we will get the same $r_c$ for both cases.  As we saw in Fig.~(2) of the main text, $r_c$ is a good order parameter for identifying the different ordered crystalline phases. Moreover, as we show in \fref{sfig2}, $r_c$ also clearly tracks the non-convexity of the energy landscape of converged eGPE solutions.  We see clearly that $r_c$ shows oscillations in the same region with $\Delta E \neq 0$ in Fig.~1(b) of the main draft indicating the `curing' of the MRI due to the LII for higher $E_0$ values. 

\begin{figure*}
	\centering
	\includegraphics[width = \textwidth]{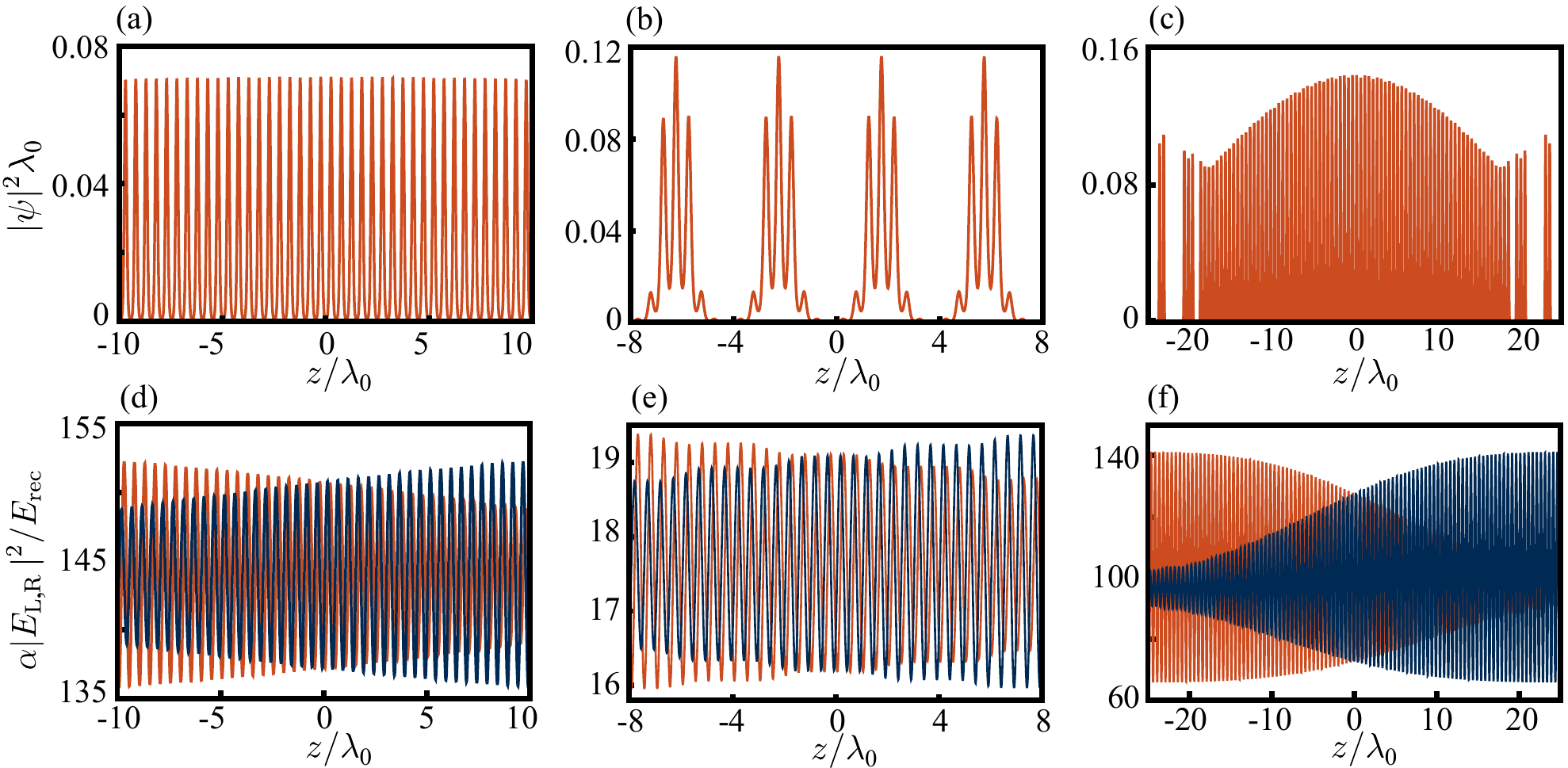}
	\caption{(a-b) Density profiles shown in Fig.~1 (d-e) of the main paper respectively. (c) Density profile corresponding to $\{g_d/g,\sqrt{\alpha}E_0/E_\text{rec}\}=\{1.7, 10\}$ with all other parameters the same as in Fig.~1 (d-e) of the main paper. The bottom row (d-f) depicts the light field intensity profiles corresponding to (a-c) respectively obtained from the solution of the Helmholtz equation. In (d-f), the curves with the decreasing (increasing) amplitude of oscillations as a function of $z$ represents $\vert E_L \vert^2$ ($\vert E_R \vert^2$).}
	\label{sfig1}
\end{figure*}

As an example of the behaviour of the electric fields, in~\fref{sfig1} we plot densities and their corresponding left and right propagating light fields corresponding to Figs.~1(d-e) from the main text and for a region with both strong LII and DDI [\fref{sfig1}(c)]. The development of a periodic potential breaking the translation symmetry of the light field intensity accompanying the development of the periodic crystalline structures for the atomic density is clearly shown. One feature to note in~\fref{sfig1} is that in general we find that the peak intensity of the standing-wave light monotonically decreases (in the direction of propagation of the applied travelling wave) in a region with an atomic density wave. Interestingly, this feature helps one to also identify gaps between atomic density waves as in~\fref{sfig1}(b,e) by noticing that the peak intensity is preserved in `atomic grating' free regions.


\section{Initial Guesses}\label{app:inan}
The different initial guesses used to obtain Fig.~2 of the main paper are either multigaussian $(\psi_\text{Gaussian})$, Tanh ($\psi_\text{Tanh}$) or Thomas-Fermi ($\psi_\text{TF}$) profiles, where
\begin{eqnarray}
	\nonumber\psi_\text{Gaussian} &=& \mathcal{A}_\text{Gaussian}\sum_{i=1}^\nu e^{-\dfrac{(z-z_i)^2}{2\sigma^2}},\\
	\nonumber \psi_\text{Tanh} &=& \mathcal{A}_\text{Tanh}\sqrt{\tanh(z+\sigma)-\tanh(z-\sigma)},\\
	\nonumber \psi_\text{TF} &=& \mathcal{A}_\text{TF}\sqrt{1-\dfrac{z^2}{\sigma^2}}.
\end{eqnarray}
The pre-factor $\mathcal{A}_j$ for $j\in[\text{Gaussian, Tanh, TF}]$ are normalization constants and $\sigma$ is proportional to the spatial widths. In case of multigaussian ansatzes we have used cases with $6\leq \nu \leq 10$ and $\sigma/\lambda_0 \sim 2$. For Tanh and Thomas-Fermi ansatzes $\sigma$ has been chosen such that $\vert \psi_\text{Tanh, TF}\vert^2$ spans the entire numerical box width.

In the LMC phase, the energy landscape consists of numerous local minima around the global minima for strong DDI. For $S$ different ansatzes listed above, solutions converge to qualitatively similar yet quantitatively different ground states with $P$ different energy values ($E_p$ with $p\in P$) where $P \leq S$. We employ the parameter $\Delta E=|E_p^\mathrm{max}-E_p^\mathrm{min}|$, the span between the maximum and minimum energy obtained from $S$ different initial guesses, to characterize the non-convexity of the energy landscape. We have used $S=6$ for Fig.~2 in the main paper.


\begin{figure*}
	\centering
	\includegraphics[width = \textwidth]{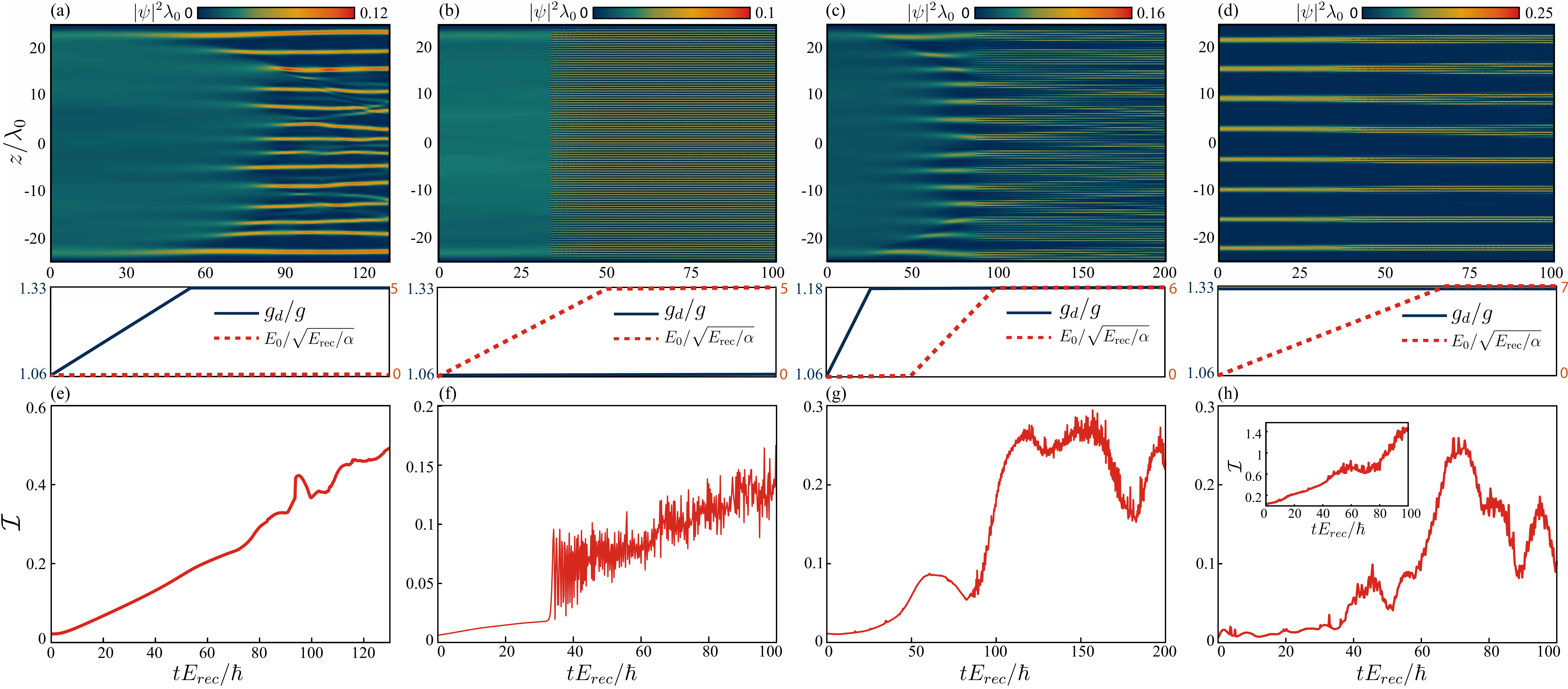}
	\caption{Preparation of (a) magnetic crystal (MC), (b) light crystal (LC), and (c) light-magnetic crystal (LMC) states starting from an unordered state with additional thermal noise, and (d) LMC starting from a MC with added thermal noise for Dy atoms with their corresponding ramping schemes of $g_d/g$ and the light intensities $E_0$ shown in the middle row, same as Fig. 3 of the main text. The bottom row (e-h) depicts the corresponding incoherence plots with time which is an average of at least three pairs of droplets where the droplets have length scales of either MC (e) or LC (f-h) . The inset in (h) is the incoherence plot when the chosen droplets are of the MC length scale.}
	\label{sfig3}
\end{figure*}
\section{Coherence Properties}\label{app:cohp}
We aim to quantify the coherence properties, following Ref.~\cite{Bland2022}, of different types of crystals generated in Fig. 3 of the main text. In Fig.~\ref{sfig3}(e)-(h) we plot the incoherence $(\mathcal{I})$ where the value zero signifies coherence and $\pi/2$ refers to incoherence. At time $t=0$ we have included noise to our initial state as follows.
\begin{eqnarray}
    \psi(z) &=& \psi_0(z) + \sum_n \alpha_n \phi_n (z)
\end{eqnarray}
where $\phi_n(z)$ are single particle states and $\alpha_n$ are complex Gaussian random variables that obey the relation,
\begin{eqnarray}
    \langle|\alpha_n|^2\rangle &=& (e^{\epsilon_n/ k_B T} - 1)^{-1} + \dfrac{1}{2}
\end{eqnarray}
We have restricted the sum to condition that $\epsilon_n \leq 2 k_B T$ with $T=10nK$. The initial noise plays an important role in the emergence of instabilities while quenching sequences. Consequently, the dynamic phase incoherence is given by,
\begin{eqnarray}
    \mathcal{I}(t) &=& \dfrac{\int_\mathcal{C} dz |\psi(z,t)|^2 [\theta(z, t) - \langle \theta(z, t)\rangle]}{\int_\mathcal{C} dz |\psi(z, t)|^2}
\end{eqnarray}
The phase is denoted by $\theta(z,t)$ and $\langle \theta(z,t)\rangle$ is chosen such that $\mathcal{I}$ is minimized at every iteration.

We evaluate incoherence between at least three pairs of droplets (of smallest length scale) and average over them for all cases. Contrasting between Fig. \ref{sfig3}(a) and (b) clearly signifies that the coherence is better maintained in the LC phase though as time progresses both monotonically lose coherence. In (c) and (d) the behaviour of coherence is extremely non monotonic but still the incoherence remains small in comparison to purely magnetic crystals. In fact, both these cases display significant recovery of coherence after the light amplitude quench and formation of light crystals.

\begin{figure*}
	\centering
	\includegraphics[width = \textwidth]{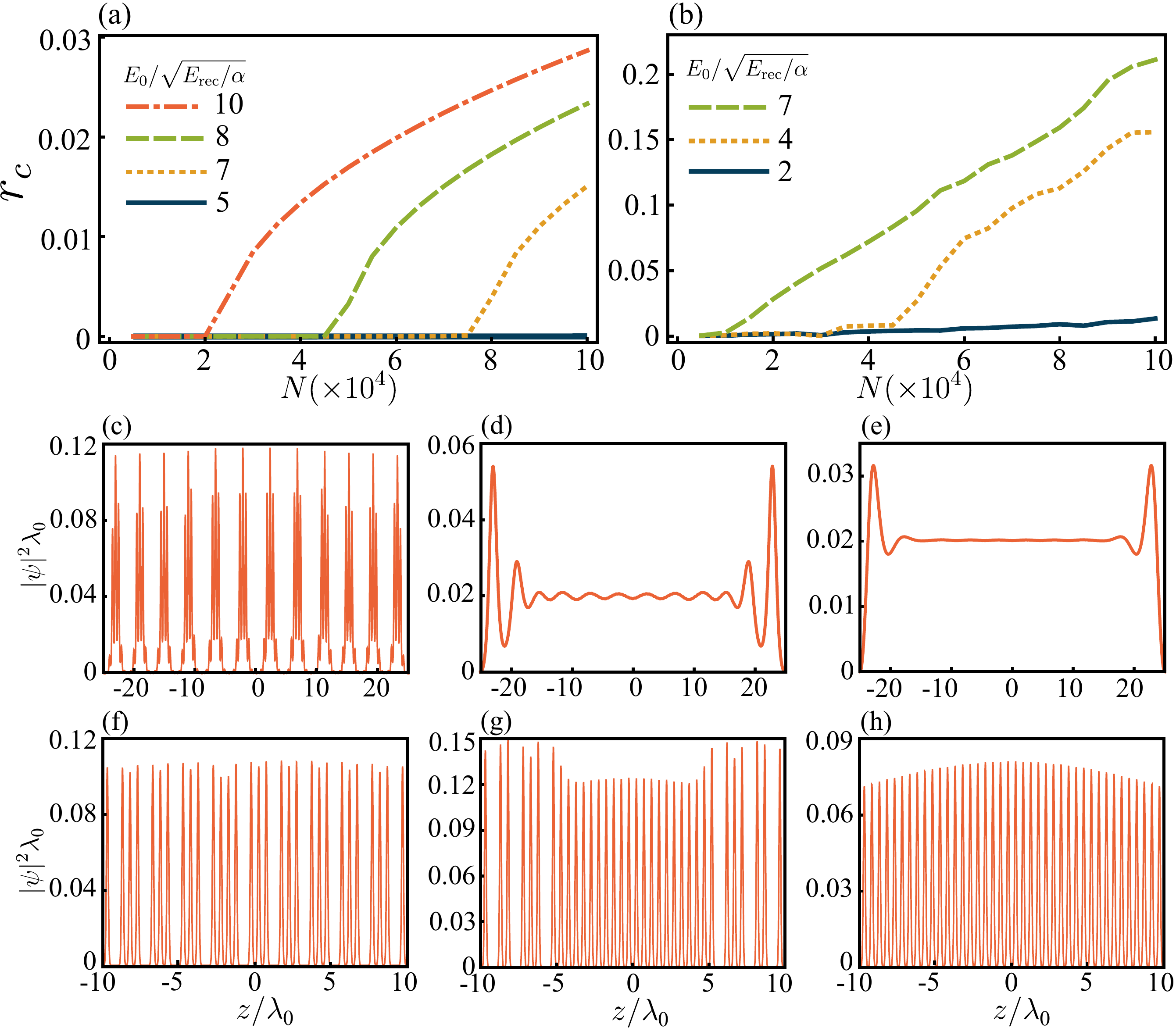}
	\caption{Demonstration of threshold value of LC captured by the parameter $r_c$ as $N$ is varied for (a) $g_d/g = 0.9$ and (b) $g_d/g = 1.88$ with other parameters similar to Fig.1 of main text. In (c)-(h) we show two cases from LMC phase where gradually reducing $N$ results in shifts of the phase boundaries and the LMC state either changing to an MC and eventually unordered state or changing into an LC state. The second row corresponds to the Fig. 1 (e) of main text except with (d) $N=80000$ and (e) $N=50000$. The third row corresponds to Fig. 1 (f) with (g) $N=50000$ and (h) $N=10000$.}
	\label{sfig4}
\end{figure*}

\section{Dependence on the Number of Atoms}\label{app:nd}
In this section we aim to have a preliminary understanding of the behaviour of the ordered phases of Fig. 2 with respect to a change in number of atoms. In case of MC phases this has been thoroughly covered both experimentally~\cite{Modugno_2019} and theoretically~\cite{Blakie_2016, Blakie_2018} in previous studies. We shall focus on the LC and LMC phases. 

The behaviour of LC with respect to number of atoms is well-captured by the $r_c$. In Fig. \ref{sfig4} the two extreme cases of $g_d/g = 0.9$ and $g_d/g = 0.1.88$ w.r.t. number of atoms are shown for different light amplitudes. As number of atoms decreases the LC is eventually lost. This is behaviour can be understood by the density dependence of the optical potential. A larger density enhances the coupling and in turn precipitates the LC phase. The loss of crystal order at lower $N$ can then compensated by increasing the light intensity. Note that, the curves in (a) are smoother compared to (b) because in the latter case, non-convexity of the energy functional makes it difficult to determine the true ground state quantitatively.

The characterization of LMC states with $N$ is, however, nontrivial. We have used the same parameters used in Fig. 1(e)-(f) of the main text to highlight the contrasting behaviour the density patterns can show when $N$ is reduced. In the second row of Fig. \ref{sfig4}, we begin with $N=10^5$ with a droplet of supersolid density pattern. When atom numbers are reduced there is a quick change in phase and the light crystalline order is lost to give rise to a density modulated state with the length scale of MC. Further reduction in $N$ results in an unordered state. On the other hand, in the third row, we begin with a state closer to the LC-LMC boundary. Here, reduction of atoms eventually results in obtaining a LC phase. This would eventually be lost to give an unordered state in accordance with the behaviour displayed in Fig. \ref{sfig4}(a)

\section{Heating Rate Calculation}\label{app:heating}
The frequency dependent polarizability of a two level atom subject to a light field with detuning $\Delta_a$ is given by [for $\Delta_a \gg \Gamma$ with $\Gamma$ denoting the linewidth (spontaneous emission rate) of the transition] \cite{steck},
\begin{align}
	\alpha(\omega) = \dfrac{\alpha_0\omega_0^2}{\omega_0^2-\omega^2} \approx \dfrac{\omega_0\alpha_0}{2\Delta_a},
\end{align}
where $\alpha_0 = 2\mu^2/\hbar\omega_0$ is the static polarizability for a transition frequency $\omega_0$ and transition electric dipole moment $\mu$. Note that the light-atom coupling parameter $\zeta = \alpha N/{2 \pi \epsilon_0\lambda_0 l^2}$ is determined by the polarizability $\alpha(\omega)$. Since in all calculations presented in the main text, we choose $\zeta = 0.1$, we will choose detuning $\Delta_a$ to ensure this is satisfied. The heating rate due to spontaneous emission for atoms subject to light of intensity $I$ is given by:
\begin{align}
	R &= \dfrac{\Gamma^3}{8\Delta_a^2}\dfrac{I}{I_{\text{sat}}},
\end{align}
with the saturation intensity $I_{\mathrm{sat}}$ given by
\begin{align}
	I_{\text{sat}} &= \dfrac{2\pi \hbar \omega_0 \Gamma}{6 \lambda_0^2}, \label{eq:satI}
\end{align}
for a two-level atom model. Focusing on Dysprosium (Dy) we find that the data for $\Gamma$ and $I_{\mathrm{sat}}$ for different optical transitions are presented in Ref.~\cite{Lev_2011}. Since the dipole moment strength $\mu$ is not directly available we estimate the same using~\eref{eq:satI} and the expression for spontaneous emission rate of a two-level atom $\Gamma = \omega_0^3 \mu^2/(3 \pi \epsilon_0 \hbar c^3)$ as:
\begin{align*}
	\mu = \sqrt{\dfrac{c\epsilon_0\hbar^2\Gamma^2}{4I_{\text{sat}}}}, 
\end{align*}
with $c$ and $\epsilon_0$ denoting the speed of light and permittivity of free space. Choosing the $\lambda_0=741$nm transition in Dy with $I_{\mathrm{sat}} = 0.57$ $\mu$W/cm$^2$ and $\Gamma = 1.12 \times 10^4$ Hz \cite{Lev_2011}, we find that a detuning of $\Delta_a \sim 2\pi \times 1.6$MHz leads to $\zeta = 0.1$ for our chosen system parameters with $N=10^5$ atoms confined in a transverse trap $\omega_\rho/2\pi=100$Hz. Note that we use the transverse trap frequency to estimate the cross-section as $l^2 \sim l_{\rho}^2 = \hbar/(m \omega_{\rho})$ in the expression of $\zeta$. This leads to the estimate of the heating rate of $R \sim 34$Hz presented in the main text.

For the sake of completeness and to show that there is enough room in terms of choice of experimental parameters, we first present below a table giving the detuning $\Delta_a$ choices and heating rates for different lines of Dy with $(N,A\equiv 2 \pi l_\rho^2)=(10^5, 4\times10^{-12} \mathrm{m}^2)$:
\begin{widetext}
\begin{center}
	\begin{tabular}{||c c c c c c c||} 
		\hline
		$\lambda_0$(nm) & $I$(W/$\text{m}^2$) & $\Gamma$(Hz) & $\mu$(Debye) & $\alpha~(\text{Hz-cm}^2/\text{V}^2)$ & $\Delta_a$(MHz) & $R$(Hz) \\ [0.5ex] 
		\hline\hline
		$1001$ & $3\times10^3$ & $330$ & 0.032 & 265$h$ & 0.5 & 20\\ 
		$741$ & $4.4\times10^3$ & $1.12\times10^4$ & 0.12 & 370$h$ & 10 & 34\\
		$626$ & $5\times10^3$ & $8.5\times10^5$ & 0.81 & 337$h$ & 500 & 52\\
		$598$ & $5\times10^3$ & $7.7\times10^4$ & 0.22 & 332$h$ & 50 & 59\\
		$421$ & $7.3\times10^3$ & $2\times10^8$ & 6.93 & 221$h$ & $5.5\times10^4$ & 113\\ [1ex] 
		\hline
	\end{tabular}
\end{center}
\end{widetext}
In a similar manner we also find the following possibilities for Erbium (Er) with $(N,A)=(5\times10^5, 4\times 10^{-12}\mathrm{m}^2)$ in all cases:
\begin{widetext}
\begin{center}
	\begin{tabular}{||c c c c c c c||} 
		\hline
		$\lambda_0$(nm) & $I$(W/$\text{m}^2$) & $\Gamma$(Hz) & $\mu$(Debye) & $\alpha~(\text{Hz-cm}^2/\text{V}^2)$ & $\Delta_a$(MHz) & $R$(Hz) \\ [0.5ex] 
		\hline\hline
		$1299$ & $1.1\times10^4$ & $5.6$ & 0.006 & 142 & 0.07$h$ & 6\\ 
		$841$ & $1.7\times10^4$ & $5\times10^4$ & 0.3 & 93 & 250$h$ & 6\\
		$631$ & $1.8\times10^4$ & $1.8\times10^5$ & 0.37 & 91 & 400$h$ & 13\\
		$582$ & $1.8\times10^4$ & $1\times10^6$ & 0.77 & 90 & 1700$h$ & 18\\
		$400$ & $1.7\times10^4$ & $1.7\times10^8$ & 5.85 & 91 & 95000 & 55\\[1ex] 
		\hline
	\end{tabular}
\end{center} 
\end{widetext}

	%
	

\begin{thebibliography}{48}%
		\makeatletter
		\providecommand \@ifxundefined [1]{%
			\@ifx{#1\undefined}
		}%
		\providecommand \@ifnum [1]{%
			\ifnum #1\expandafter \@firstoftwo
			\else \expandafter \@secondoftwo
			\fi
		}%
		\providecommand \@ifx [1]{%
			\ifx #1\expandafter \@firstoftwo
			\else \expandafter \@secondoftwo
			\fi
		}%
		\providecommand \natexlab [1]{#1}%
		\providecommand \enquote  [1]{``#1''}%
		\providecommand \bibnamefont  [1]{#1}%
		\providecommand \bibfnamefont [1]{#1}%
		\providecommand \citenamefont [1]{#1}%
		\providecommand \href@noop [0]{\@secondoftwo}%
		\providecommand \href [0]{\begingroup \@sanitize@url \@href}%
		\providecommand \@href[1]{\@@startlink{#1}\@@href}%
		\providecommand \@@href[1]{\endgroup#1\@@endlink}%
		\providecommand \@sanitize@url [0]{\catcode `\\12\catcode `\$12\catcode
			`\&12\catcode `\#12\catcode `\^12\catcode `\_12\catcode `\%12\relax}%
		\providecommand \@@startlink[1]{}%
		\providecommand \@@endlink[0]{}%
		\providecommand \url  [0]{\begingroup\@sanitize@url \@url }%
		\providecommand \@url [1]{\endgroup\@href {#1}{\urlprefix }}%
		\providecommand \urlprefix  [0]{URL }%
		\providecommand \Eprint [0]{\href }%
		\providecommand \doibase [0]{http://dx.doi.org/}%
		\providecommand \selectlanguage [0]{\@gobble}%
		\providecommand \bibinfo  [0]{\@secondoftwo}%
		\providecommand \bibfield  [0]{\@secondoftwo}%
		\providecommand \translation [1]{[#1]}%
		\providecommand \BibitemOpen [0]{}%
		\providecommand \bibitemStop [0]{}%
		\providecommand \bibitemNoStop [0]{.\EOS\space}%
		\providecommand \EOS [0]{\spacefactor3000\relax}%
		\providecommand \BibitemShut  [1]{\csname bibitem#1\endcsname}%
		\let\auto@bib@innerbib\@empty
		\bibitem [{\citenamefont {Defenu}\ \emph {et~al.}(2021)\citenamefont {Defenu},
			\citenamefont {Donner}, \citenamefont {Macrì}, \citenamefont {Pagano},
			\citenamefont {Ruffo},\ and\ \citenamefont
			{Trombettoni}}]{defenu_long-range_2021}%
		\BibitemOpen
		\bibfield  {author} {\bibinfo {author} {\bibfnamefont {N.}~\bibnamefont
				{Defenu}}, \bibinfo {author} {\bibfnamefont {T.}~\bibnamefont {Donner}},
			\bibinfo {author} {\bibfnamefont {T.}~\bibnamefont {Macrì}}, \bibinfo
			{author} {\bibfnamefont {G.}~\bibnamefont {Pagano}}, \bibinfo {author}
			{\bibfnamefont {S.}~\bibnamefont {Ruffo}}, \ and\ \bibinfo {author}
			{\bibfnamefont {A.}~\bibnamefont {Trombettoni}},\ }\href@noop {} {\
			(\bibinfo {year} {2021})},\ \Eprint {http://arxiv.org/abs/arXiv:2109.01063}
		{arXiv:2109.01063} \BibitemShut {NoStop}%
		\bibitem [{\citenamefont {Santos}\ \emph {et~al.}(2000)\citenamefont {Santos},
			\citenamefont {Shlyapnikov}, \citenamefont {Zoller},\ and\ \citenamefont
			{Lewenstein}}]{Lewenstein_2000}%
		\BibitemOpen
		\bibfield  {author} {\bibinfo {author} {\bibfnamefont {L.}~\bibnamefont
				{Santos}}, \bibinfo {author} {\bibfnamefont {G.~V.}\ \bibnamefont
				{Shlyapnikov}}, \bibinfo {author} {\bibfnamefont {P.}~\bibnamefont {Zoller}},
			\ and\ \bibinfo {author} {\bibfnamefont {M.}~\bibnamefont {Lewenstein}},\
		}\href {\doibase 10.1103/PhysRevLett.85.1791} {\bibfield  {journal} {\bibinfo
				{journal} {Phys. Rev. Lett.}\ }\textbf {\bibinfo {volume} {85}},\ \bibinfo
			{pages} {1791} (\bibinfo {year} {2000})}\BibitemShut {NoStop}%
		\bibitem [{\citenamefont {Lahaye}\ \emph {et~al.}(2009)\citenamefont {Lahaye},
			\citenamefont {Menotti}, \citenamefont {Santos}, \citenamefont {Lewenstein},\
			and\ \citenamefont {Pfau}}]{lahaye_physics_2009}%
		\BibitemOpen
		\bibfield  {author} {\bibinfo {author} {\bibfnamefont {T.}~\bibnamefont
				{Lahaye}}, \bibinfo {author} {\bibfnamefont {C.}~\bibnamefont {Menotti}},
			\bibinfo {author} {\bibfnamefont {L.}~\bibnamefont {Santos}}, \bibinfo
			{author} {\bibfnamefont {M.}~\bibnamefont {Lewenstein}}, \ and\ \bibinfo
			{author} {\bibfnamefont {T.}~\bibnamefont {Pfau}},\ }\href {\doibase
			10.1088/0034-4885/72/12/126401} {\bibfield  {journal} {\bibinfo  {journal}
				{Reports on Progress in Physics}\ }\textbf {\bibinfo {volume} {72}},\
			\bibinfo {pages} {126401} (\bibinfo {year} {2009})}\BibitemShut {NoStop}%
		\bibitem [{\citenamefont {Baranov}\ \emph {et~al.}(2012)\citenamefont
			{Baranov}, \citenamefont {Dalmonte}, \citenamefont {Pupillo},\ and\
			\citenamefont {Zoller}}]{Baranov2012}%
		\BibitemOpen
		\bibfield  {author} {\bibinfo {author} {\bibfnamefont {M.~A.}\ \bibnamefont
				{Baranov}}, \bibinfo {author} {\bibfnamefont {M.}~\bibnamefont {Dalmonte}},
			\bibinfo {author} {\bibfnamefont {G.}~\bibnamefont {Pupillo}}, \ and\
			\bibinfo {author} {\bibfnamefont {P.}~\bibnamefont {Zoller}},\ }\href
		{\doibase 10.1021/cr2003568} {\bibfield  {journal} {\bibinfo  {journal}
				{Chemical Reviews}\ }\textbf {\bibinfo {volume} {112}},\ \bibinfo {pages}
			{5012} (\bibinfo {year} {2012})}\BibitemShut {NoStop}%
		\bibitem [{\citenamefont {Norcia}\ and\ \citenamefont
			{Ferlaino}(2021)}]{norcia_developments_2021}%
		\BibitemOpen
		\bibfield  {author} {\bibinfo {author} {\bibfnamefont {M.~A.}\ \bibnamefont
				{Norcia}}\ and\ \bibinfo {author} {\bibfnamefont {F.}~\bibnamefont
				{Ferlaino}},\ }\href {\doibase 10.1038/s41567-021-01398-7} {\bibfield
			{journal} {\bibinfo  {journal} {Nature Physics}\ }\textbf {\bibinfo {volume}
				{17}},\ \bibinfo {pages} {1349} (\bibinfo {year} {2021})}\BibitemShut
		{NoStop}%
		\bibitem [{\citenamefont {Chomaz}\ \emph {et~al.}(2022)\citenamefont {Chomaz},
			\citenamefont {Ferrier-Barbut}, \citenamefont {Ferlaino}, \citenamefont
			{Laburthe-Tolra}, \citenamefont {Lev},\ and\ \citenamefont
			{Pfau}}]{chomaz_dipolar_2022}%
		\BibitemOpen
		\bibfield  {author} {\bibinfo {author} {\bibfnamefont {L.}~\bibnamefont
				{Chomaz}}, \bibinfo {author} {\bibfnamefont {I.}~\bibnamefont
				{Ferrier-Barbut}}, \bibinfo {author} {\bibfnamefont {F.}~\bibnamefont
				{Ferlaino}}, \bibinfo {author} {\bibfnamefont {B.}~\bibnamefont
				{Laburthe-Tolra}}, \bibinfo {author} {\bibfnamefont {B.~L.}\ \bibnamefont
				{Lev}}, \ and\ \bibinfo {author} {\bibfnamefont {T.}~\bibnamefont {Pfau}},\
		}\href@noop {} {\  (\bibinfo {year} {2022})},\ \Eprint
		{http://arxiv.org/abs/arXiv:2201.02672} {arXiv:2201.02672} \BibitemShut
		{NoStop}%
		\bibitem [{\citenamefont {Giovanazzi}\ \emph {et~al.}(2002)\citenamefont
			{Giovanazzi}, \citenamefont {O'Dell},\ and\ \citenamefont
			{Kurizki}}]{Kurizki_2002}%
		\BibitemOpen
		\bibfield  {author} {\bibinfo {author} {\bibfnamefont {S.}~\bibnamefont
				{Giovanazzi}}, \bibinfo {author} {\bibfnamefont {D.}~\bibnamefont {O'Dell}},
			\ and\ \bibinfo {author} {\bibfnamefont {G.}~\bibnamefont {Kurizki}},\ }\href
		{\doibase 10.1103/PhysRevLett.88.130402} {\bibfield  {journal} {\bibinfo
				{journal} {Phys. Rev. Lett.}\ }\textbf {\bibinfo {volume} {88}},\ \bibinfo
			{pages} {130402} (\bibinfo {year} {2002})}\BibitemShut {NoStop}%
		\bibitem [{\citenamefont {O'Dell}\ \emph {et~al.}(2003)\citenamefont {O'Dell},
			\citenamefont {Giovanazzi},\ and\ \citenamefont {Kurizki}}]{Kurziki_2003}%
		\BibitemOpen
		\bibfield  {author} {\bibinfo {author} {\bibfnamefont {D.~H.~J.}\
				\bibnamefont {O'Dell}}, \bibinfo {author} {\bibfnamefont {S.}~\bibnamefont
				{Giovanazzi}}, \ and\ \bibinfo {author} {\bibfnamefont {G.}~\bibnamefont
				{Kurizki}},\ }\href {\doibase 10.1103/PhysRevLett.90.110402} {\bibfield
			{journal} {\bibinfo  {journal} {Phys. Rev. Lett.}\ }\textbf {\bibinfo
				{volume} {90}},\ \bibinfo {pages} {110402} (\bibinfo {year}
			{2003})}\BibitemShut {NoStop}%
		\bibitem [{\citenamefont {Honer}\ \emph {et~al.}(2010)\citenamefont {Honer},
			\citenamefont {Weimer}, \citenamefont {Pfau},\ and\ \citenamefont
			{B\"uchler}}]{honer_collective_2010}%
		\BibitemOpen
		\bibfield  {author} {\bibinfo {author} {\bibfnamefont {J.}~\bibnamefont
				{Honer}}, \bibinfo {author} {\bibfnamefont {H.}~\bibnamefont {Weimer}},
			\bibinfo {author} {\bibfnamefont {T.}~\bibnamefont {Pfau}}, \ and\ \bibinfo
			{author} {\bibfnamefont {H.~P.}\ \bibnamefont {B\"uchler}},\ }\href {\doibase
			10.1103/PhysRevLett.105.160404} {\bibfield  {journal} {\bibinfo  {journal}
				{Phys. Rev. Lett.}\ }\textbf {\bibinfo {volume} {105}},\ \bibinfo {pages}
			{160404} (\bibinfo {year} {2010})}\BibitemShut {NoStop}%
		\bibitem [{\citenamefont {Ostermann}\ \emph {et~al.}(2016)\citenamefont
			{Ostermann}, \citenamefont {Piazza},\ and\ \citenamefont
			{Ritsch}}]{Ritsch_2016}%
		\BibitemOpen
		\bibfield  {author} {\bibinfo {author} {\bibfnamefont {S.}~\bibnamefont
				{Ostermann}}, \bibinfo {author} {\bibfnamefont {F.}~\bibnamefont {Piazza}}, \
			and\ \bibinfo {author} {\bibfnamefont {H.}~\bibnamefont {Ritsch}},\ }\href
		{\doibase 10.1103/PhysRevX.6.021026} {\bibfield  {journal} {\bibinfo
				{journal} {Phys. Rev. X}\ }\textbf {\bibinfo {volume} {6}},\ \bibinfo {pages}
			{021026} (\bibinfo {year} {2016})}\BibitemShut {NoStop}%
		\bibitem [{\citenamefont {Ostermann}\ \emph {et~al.}(2017)\citenamefont
			{Ostermann}, \citenamefont {Piazza},\ and\ \citenamefont
			{Ritsch}}]{ostermann_probing_2017}%
		\BibitemOpen
		\bibfield  {author} {\bibinfo {author} {\bibfnamefont {S.}~\bibnamefont
				{Ostermann}}, \bibinfo {author} {\bibfnamefont {F.}~\bibnamefont {Piazza}}, \
			and\ \bibinfo {author} {\bibfnamefont {H.}~\bibnamefont {Ritsch}},\ }\href
		{\doibase 10.1088/1367-2630/aa91c3} {\bibfield  {journal} {\bibinfo
				{journal} {New Journal of Physics}\ }\textbf {\bibinfo {volume} {19}},\
			\bibinfo {pages} {125002} (\bibinfo {year} {2017})}\BibitemShut {NoStop}%
		\bibitem [{\citenamefont {Dimitrova}\ \emph {et~al.}(2017)\citenamefont
			{Dimitrova}, \citenamefont {Lunden}, \citenamefont {Amato-Grill},
			\citenamefont {Jepsen}, \citenamefont {Yu}, \citenamefont {Messer},
			\citenamefont {Rigaldo}, \citenamefont {Puentes}, \citenamefont {Weld},\ and\
			\citenamefont {Ketterle}}]{dimitrova_observation_2017}%
		\BibitemOpen
		\bibfield  {author} {\bibinfo {author} {\bibfnamefont {I.}~\bibnamefont
				{Dimitrova}}, \bibinfo {author} {\bibfnamefont {W.}~\bibnamefont {Lunden}},
			\bibinfo {author} {\bibfnamefont {J.}~\bibnamefont {Amato-Grill}}, \bibinfo
			{author} {\bibfnamefont {N.}~\bibnamefont {Jepsen}}, \bibinfo {author}
			{\bibfnamefont {Y.}~\bibnamefont {Yu}}, \bibinfo {author} {\bibfnamefont
				{M.}~\bibnamefont {Messer}}, \bibinfo {author} {\bibfnamefont
				{T.}~\bibnamefont {Rigaldo}}, \bibinfo {author} {\bibfnamefont
				{G.}~\bibnamefont {Puentes}}, \bibinfo {author} {\bibfnamefont
				{D.}~\bibnamefont {Weld}}, \ and\ \bibinfo {author} {\bibfnamefont
				{W.}~\bibnamefont {Ketterle}},\ }\href {\doibase 10.1103/PhysRevA.96.051603}
		{\bibfield  {journal} {\bibinfo  {journal} {Phys. Rev. A}\ }\textbf {\bibinfo
				{volume} {96}},\ \bibinfo {pages} {051603} (\bibinfo {year}
			{2017})}\BibitemShut {NoStop}%
		\bibitem [{\citenamefont {Zhang}\ \emph {et~al.}(2018)\citenamefont {Zhang},
  \citenamefont {Walther},\ and\ \citenamefont {Pohl}}]{Zhang2018}%
  \BibitemOpen
  \bibfield  {author} {\bibinfo {author} {\bibfnamefont {Y.-C.}\ \bibnamefont
  {Zhang}}, \bibinfo {author} {\bibfnamefont {V.}~\bibnamefont {Walther}}, \
  and\ \bibinfo {author} {\bibfnamefont {T.}~\bibnamefont {Pohl}},\ }\href
  {\doibase 10.1103/PhysRevLett.121.073604} {\bibfield  {journal} {\bibinfo
  {journal} {Physical Review Letters}\ }\textbf {\bibinfo {volume} {121}},\
  \bibinfo {pages} {073604} (\bibinfo {year} {2018})},\ \Eprint
  {http://arxiv.org/abs/1805.03422} {arXiv:1805.03422} \BibitemShut {NoStop}%
\bibitem [{\citenamefont {Zhang}\ \emph {et~al.}(2021)\citenamefont {Zhang},
  \citenamefont {Walther},\ and\ \citenamefont {Pohl}}]{Zhang2021}%
  \BibitemOpen
  \bibfield  {author} {\bibinfo {author} {\bibfnamefont {Y.-C.}\ \bibnamefont
  {Zhang}}, \bibinfo {author} {\bibfnamefont {V.}~\bibnamefont {Walther}}, \
  and\ \bibinfo {author} {\bibfnamefont {T.}~\bibnamefont {Pohl}},\ }\href
  {\doibase 10.1103/PhysRevA.103.023308} {\bibfield  {journal} {\bibinfo
  {journal} {Physical Review A}\ }\textbf {\bibinfo {volume} {103}},\ \bibinfo
  {pages} {023308} (\bibinfo {year} {2021})},\ \Eprint
  {http://arxiv.org/abs/2011.04615} {arXiv:2011.04615} \BibitemShut {NoStop}%
   \bibitem [{\citenamefont {Chatterjee}\ and\ \citenamefont
			{Lode}(2018)}]{Chatterjee2018}%
		\BibitemOpen
		\bibfield  {author} {\bibinfo {author} {\bibfnamefont {B.}~\bibnamefont
				{Chatterjee}}\ and\ \bibinfo {author} {\bibfnamefont {A.~U.~J.}\ \bibnamefont
				{Lode}},\ }\href {\doibase 10.1103/PhysRevA.98.053624} {\bibfield  {journal}
			{\bibinfo  {journal} {Phys. Rev. A}\ }\textbf {\bibinfo {volume} {98}},\
			\bibinfo {pages} {053624} (\bibinfo {year} {2018})}\BibitemShut {NoStop}%
		\bibitem [{\citenamefont {Chatterjee}\ \emph {et~al.}(2020)\citenamefont
			{Chatterjee}, \citenamefont {L\'ev\^eque}, \citenamefont {Schmiedmayer},\
			and\ \citenamefont {Lode}}]{Chatterjee2022}%
		\BibitemOpen
		\bibfield  {author} {\bibinfo {author} {\bibfnamefont {B.}~\bibnamefont
				{Chatterjee}}, \bibinfo {author} {\bibfnamefont {C.}~\bibnamefont
				{L\'ev\^eque}}, \bibinfo {author} {\bibfnamefont {J.}~\bibnamefont
				{Schmiedmayer}}, \ and\ \bibinfo {author} {\bibfnamefont {A.~U.~J.}\
				\bibnamefont {Lode}},\ }\href {\doibase 10.1103/PhysRevLett.125.093602}
		{\bibfield  {journal} {\bibinfo  {journal} {Phys. Rev. Lett.}\ }\textbf
			{\bibinfo {volume} {125}},\ \bibinfo {pages} {093602} (\bibinfo {year}
			{2020})}\BibitemShut {NoStop}%
		\bibitem [{\citenamefont {Mottl}\ \emph {et~al.}(2012)\citenamefont {Mottl},
			\citenamefont {Brennecke}, \citenamefont {Baumann}, \citenamefont {Landig},
			\citenamefont {Donner},\ and\ \citenamefont {Esslinger}}]{Esslinger_2012}%
		\BibitemOpen
		\bibfield  {author} {\bibinfo {author} {\bibfnamefont {R.}~\bibnamefont
				{Mottl}}, \bibinfo {author} {\bibfnamefont {F.}~\bibnamefont {Brennecke}},
			\bibinfo {author} {\bibfnamefont {K.}~\bibnamefont {Baumann}}, \bibinfo
			{author} {\bibfnamefont {R.}~\bibnamefont {Landig}}, \bibinfo {author}
			{\bibfnamefont {T.}~\bibnamefont {Donner}}, \ and\ \bibinfo {author}
			{\bibfnamefont {T.}~\bibnamefont {Esslinger}},\ }\href {\doibase
			10.1126/science.1220314} {\bibfield  {journal} {\bibinfo  {journal}
				{Science}\ }\textbf {\bibinfo {volume} {336}},\ \bibinfo {pages} {1570}
			(\bibinfo {year} {2012})}\BibitemShut {NoStop}%
		\bibitem [{\citenamefont {Ritsch}\ \emph {et~al.}(2013)\citenamefont {Ritsch},
			\citenamefont {Domokos}, \citenamefont {Brennecke},\ and\ \citenamefont
			{Esslinger}}]{ritsch_cold_2013}%
		\BibitemOpen
		\bibfield  {author} {\bibinfo {author} {\bibfnamefont {H.}~\bibnamefont
				{Ritsch}}, \bibinfo {author} {\bibfnamefont {P.}~\bibnamefont {Domokos}},
			\bibinfo {author} {\bibfnamefont {F.}~\bibnamefont {Brennecke}}, \ and\
			\bibinfo {author} {\bibfnamefont {T.}~\bibnamefont {Esslinger}},\ }\href
		{\doibase 10.1103/RevModPhys.85.553} {\bibfield  {journal} {\bibinfo
				{journal} {Rev. Mod. Phys.}\ }\textbf {\bibinfo {volume} {85}},\ \bibinfo
			{pages} {553} (\bibinfo {year} {2013})}\BibitemShut {NoStop}%
		\bibitem [{\citenamefont {Mivehvar}\ \emph {et~al.}(2021)\citenamefont
			{Mivehvar}, \citenamefont {Piazza}, \citenamefont {Donner},\ and\
			\citenamefont {Ritsch}}]{Mivehvar2021Cavity}%
		\BibitemOpen
		\bibfield  {author} {\bibinfo {author} {\bibfnamefont {F.}~\bibnamefont
				{Mivehvar}}, \bibinfo {author} {\bibfnamefont {F.}~\bibnamefont {Piazza}},
			\bibinfo {author} {\bibfnamefont {T.}~\bibnamefont {Donner}}, \ and\ \bibinfo
			{author} {\bibfnamefont {H.}~\bibnamefont {Ritsch}},\ }\href {\doibase
			10.1080/00018732.2021.1969727} {\bibfield  {journal} {\bibinfo  {journal}
				{Advances in Physics}\ }\textbf {\bibinfo {volume} {70}},\ \bibinfo {pages}
			{1} (\bibinfo {year} {2021})}\BibitemShut {NoStop}%
		\bibitem [{\citenamefont {Karpov}\ and\ \citenamefont
			{Piazza}(2022)}]{Karpov2022}%
		\BibitemOpen
		\bibfield  {author} {\bibinfo {author} {\bibfnamefont {P.}~\bibnamefont
				{Karpov}}\ and\ \bibinfo {author} {\bibfnamefont {F.}~\bibnamefont
				{Piazza}},\ }\href {\doibase 10.1103/PhysRevLett.128.103201} {\bibfield
			{journal} {\bibinfo  {journal} {Phys. Rev. Lett.}\ }\textbf {\bibinfo
				{volume} {128}},\ \bibinfo {pages} {103201} (\bibinfo {year}
			{2022})}\BibitemShut {NoStop}%
		\bibitem [{\citenamefont {Gopalakrishnan}\ \emph {et~al.}(2009)\citenamefont
			{Gopalakrishnan}, \citenamefont {Lev},\ and\ \citenamefont
			{Goldbart}}]{Lev1}%
		\BibitemOpen
		\bibfield  {author} {\bibinfo {author} {\bibfnamefont {S.}~\bibnamefont
				{Gopalakrishnan}}, \bibinfo {author} {\bibfnamefont {B.~L.}\ \bibnamefont
				{Lev}}, \ and\ \bibinfo {author} {\bibfnamefont {P.~M.}\ \bibnamefont
				{Goldbart}},\ }\href {\doibase 10.1038/nphys1403} {\bibfield  {journal}
			{\bibinfo  {journal} {Nature Physics}\ }\textbf {\bibinfo {volume} {5}},\
			\bibinfo {pages} {845} (\bibinfo {year} {2009})}\BibitemShut {NoStop}%
		\bibitem [{\citenamefont {Koll{\'{a}}r}\ \emph {et~al.}(2017)\citenamefont
			{Koll{\'{a}}r}, \citenamefont {Papageorge}, \citenamefont {Vaidya},
			\citenamefont {Guo}, \citenamefont {Keeling},\ and\ \citenamefont
			{Lev}}]{Lev2}%
		\BibitemOpen
		\bibfield  {author} {\bibinfo {author} {\bibfnamefont {A.~J.}\ \bibnamefont
				{Koll{\'{a}}r}}, \bibinfo {author} {\bibfnamefont {A.~T.}\ \bibnamefont
				{Papageorge}}, \bibinfo {author} {\bibfnamefont {V.~D.}\ \bibnamefont
				{Vaidya}}, \bibinfo {author} {\bibfnamefont {Y.}~\bibnamefont {Guo}},
			\bibinfo {author} {\bibfnamefont {J.}~\bibnamefont {Keeling}}, \ and\
			\bibinfo {author} {\bibfnamefont {B.~L.}\ \bibnamefont {Lev}},\ }\href
		{\doibase 10.1038/ncomms14386} {\bibfield  {journal} {\bibinfo  {journal}
				{Nature Communications}\ }\textbf {\bibinfo {volume} {8}},\ \bibinfo {pages}
			{14386} (\bibinfo {year} {2017})}\BibitemShut {NoStop}%
		\bibitem [{\citenamefont {Vaidya}\ \emph {et~al.}(2018)\citenamefont {Vaidya},
			\citenamefont {Guo}, \citenamefont {Kroeze}, \citenamefont {Ballantine},
			\citenamefont {Koll\'ar}, \citenamefont {Keeling},\ and\ \citenamefont
			{Lev}}]{Lev3}%
		\BibitemOpen
		\bibfield  {author} {\bibinfo {author} {\bibfnamefont {V.~D.}\ \bibnamefont
				{Vaidya}}, \bibinfo {author} {\bibfnamefont {Y.}~\bibnamefont {Guo}},
			\bibinfo {author} {\bibfnamefont {R.~M.}\ \bibnamefont {Kroeze}}, \bibinfo
			{author} {\bibfnamefont {K.~E.}\ \bibnamefont {Ballantine}}, \bibinfo
			{author} {\bibfnamefont {A.~J.}\ \bibnamefont {Koll\'ar}}, \bibinfo {author}
			{\bibfnamefont {J.}~\bibnamefont {Keeling}}, \ and\ \bibinfo {author}
			{\bibfnamefont {B.~L.}\ \bibnamefont {Lev}},\ }\href {\doibase
			10.1103/PhysRevX.8.011002} {\bibfield  {journal} {\bibinfo  {journal} {Phys.
					Rev. X}\ }\textbf {\bibinfo {volume} {8}},\ \bibinfo {pages} {011002}
			(\bibinfo {year} {2018})}\BibitemShut {NoStop}%
		\bibitem [{\citenamefont {Deng}\ \emph {et~al.}(2012)\citenamefont {Deng},
			\citenamefont {Cheng}, \citenamefont {Jing}, \citenamefont {Sun},\ and\
			\citenamefont {Yi}}]{Deng2012}%
		\BibitemOpen
		\bibfield  {author} {\bibinfo {author} {\bibfnamefont {Y.}~\bibnamefont
				{Deng}}, \bibinfo {author} {\bibfnamefont {J.}~\bibnamefont {Cheng}},
			\bibinfo {author} {\bibfnamefont {H.}~\bibnamefont {Jing}}, \bibinfo {author}
			{\bibfnamefont {C.-P.}\ \bibnamefont {Sun}}, \ and\ \bibinfo {author}
			{\bibfnamefont {S.}~\bibnamefont {Yi}},\ }\href {\doibase
			10.1103/PhysRevLett.108.125301} {\bibfield  {journal} {\bibinfo  {journal}
				{Phys. Rev. Lett.}\ }\textbf {\bibinfo {volume} {108}},\ \bibinfo {pages}
			{125301} (\bibinfo {year} {2012})}\BibitemShut {NoStop}%
		\bibitem [{\citenamefont {Gopalakrishnan}\ \emph {et~al.}(2013)\citenamefont
			{Gopalakrishnan}, \citenamefont {Martin},\ and\ \citenamefont
			{Demler}}]{Sarang2013}%
		\BibitemOpen
		\bibfield  {author} {\bibinfo {author} {\bibfnamefont {S.}~\bibnamefont
				{Gopalakrishnan}}, \bibinfo {author} {\bibfnamefont {I.}~\bibnamefont
				{Martin}}, \ and\ \bibinfo {author} {\bibfnamefont {E.~A.}\ \bibnamefont
				{Demler}},\ }\href {\doibase 10.1103/PhysRevLett.111.185304} {\bibfield
			{journal} {\bibinfo  {journal} {Phys. Rev. Lett.}\ }\textbf {\bibinfo
				{volume} {111}},\ \bibinfo {pages} {185304} (\bibinfo {year}
			{2013})}\BibitemShut {NoStop}%
		\bibitem [{\citenamefont {Li}\ \emph {et~al.}(2019)\citenamefont {Li},
			\citenamefont {Wang}, \citenamefont {Wang}, \citenamefont {Shi},
			\citenamefont {Jardine},\ and\ \citenamefont {Wen}}]{Li_2019}%
		\BibitemOpen
		\bibfield  {author} {\bibinfo {author} {\bibfnamefont {X.}~\bibnamefont
				{Li}}, \bibinfo {author} {\bibfnamefont {Q.}~\bibnamefont {Wang}}, \bibinfo
			{author} {\bibfnamefont {H.}~\bibnamefont {Wang}}, \bibinfo {author}
			{\bibfnamefont {C.}~\bibnamefont {Shi}}, \bibinfo {author} {\bibfnamefont
				{M.}~\bibnamefont {Jardine}}, \ and\ \bibinfo {author} {\bibfnamefont
				{L.}~\bibnamefont {Wen}},\ }\href {\doibase 10.1088/1361-6455/ab2a9b}
		{\bibfield  {journal} {\bibinfo  {journal} {Journal of Physics B: Atomic,
					Molecular and Optical Physics}\ }\textbf {\bibinfo {volume} {52}},\ \bibinfo
			{pages} {155302} (\bibinfo {year} {2019})}\BibitemShut {NoStop}%
		\bibitem [{\citenamefont {Mivehvar}\ \emph {et~al.}(2019)\citenamefont
			{Mivehvar}, \citenamefont {Ritsch},\ and\ \citenamefont
			{Piazza}}]{Mivehvar2019}%
		\BibitemOpen
		\bibfield  {author} {\bibinfo {author} {\bibfnamefont {F.}~\bibnamefont
				{Mivehvar}}, \bibinfo {author} {\bibfnamefont {H.}~\bibnamefont {Ritsch}}, \
			and\ \bibinfo {author} {\bibfnamefont {F.}~\bibnamefont {Piazza}},\ }\href
		{\doibase 10.1103/PhysRevLett.123.210604} {\bibfield  {journal} {\bibinfo
				{journal} {Phys. Rev. Lett.}\ }\textbf {\bibinfo {volume} {123}},\ \bibinfo
			{pages} {210604} (\bibinfo {year} {2019})}\BibitemShut {NoStop}%
		\bibitem [{\citenamefont {Chomaz}\ \emph {et~al.}(2019)\citenamefont {Chomaz},
			\citenamefont {Petter}, \citenamefont {Ilzh\"ofer}, \citenamefont {Natale},
			\citenamefont {Trautmann}, \citenamefont {Politi}, \citenamefont
			{Durastante}, \citenamefont {van Bijnen}, \citenamefont {Patscheider},
			\citenamefont {Sohmen}, \citenamefont {Mark},\ and\ \citenamefont
			{Ferlaino}}]{Ferlaino_2019}%
		\BibitemOpen
		\bibfield  {author} {\bibinfo {author} {\bibfnamefont {L.}~\bibnamefont
				{Chomaz}}, \bibinfo {author} {\bibfnamefont {D.}~\bibnamefont {Petter}},
			\bibinfo {author} {\bibfnamefont {P.}~\bibnamefont {Ilzh\"ofer}}, \bibinfo
			{author} {\bibfnamefont {G.}~\bibnamefont {Natale}}, \bibinfo {author}
			{\bibfnamefont {A.}~\bibnamefont {Trautmann}}, \bibinfo {author}
			{\bibfnamefont {C.}~\bibnamefont {Politi}}, \bibinfo {author} {\bibfnamefont
				{G.}~\bibnamefont {Durastante}}, \bibinfo {author} {\bibfnamefont {R.~M.~W.}\
				\bibnamefont {van Bijnen}}, \bibinfo {author} {\bibfnamefont
				{A.}~\bibnamefont {Patscheider}}, \bibinfo {author} {\bibfnamefont
				{M.}~\bibnamefont {Sohmen}}, \bibinfo {author} {\bibfnamefont {M.~J.}\
				\bibnamefont {Mark}}, \ and\ \bibinfo {author} {\bibfnamefont
				{F.}~\bibnamefont {Ferlaino}},\ }\href {\doibase 10.1103/PhysRevX.9.021012}
		{\bibfield  {journal} {\bibinfo  {journal} {Phys. Rev. X}\ }\textbf {\bibinfo
				{volume} {9}},\ \bibinfo {pages} {021012} (\bibinfo {year}
			{2019})}\BibitemShut {NoStop}%
		\bibitem [{\citenamefont {B\"ottcher}\ \emph {et~al.}(2019)\citenamefont
			{B\"ottcher}, \citenamefont {Schmidt}, \citenamefont {Wenzel}, \citenamefont
			{Hertkorn}, \citenamefont {Guo}, \citenamefont {Langen},\ and\ \citenamefont
			{Pfau}}]{bottcher_transient_2019}%
		\BibitemOpen
		\bibfield  {author} {\bibinfo {author} {\bibfnamefont {F.}~\bibnamefont
				{B\"ottcher}}, \bibinfo {author} {\bibfnamefont {J.-N.}\ \bibnamefont
				{Schmidt}}, \bibinfo {author} {\bibfnamefont {M.}~\bibnamefont {Wenzel}},
			\bibinfo {author} {\bibfnamefont {J.}~\bibnamefont {Hertkorn}}, \bibinfo
			{author} {\bibfnamefont {M.}~\bibnamefont {Guo}}, \bibinfo {author}
			{\bibfnamefont {T.}~\bibnamefont {Langen}}, \ and\ \bibinfo {author}
			{\bibfnamefont {T.}~\bibnamefont {Pfau}},\ }\href {\doibase
			10.1103/PhysRevX.9.011051} {\bibfield  {journal} {\bibinfo  {journal} {Phys.
					Rev. X}\ }\textbf {\bibinfo {volume} {9}},\ \bibinfo {pages} {011051}
			(\bibinfo {year} {2019})}\BibitemShut {NoStop}%
		\bibitem [{\citenamefont {Guo}\ \emph {et~al.}(2019)\citenamefont {Guo},
			\citenamefont {B{\"o}ttcher}, \citenamefont {Hertkorn}, \citenamefont
			{Schmidt}, \citenamefont {Wenzel}, \citenamefont {B{\"u}chler}, \citenamefont
			{Langen},\ and\ \citenamefont {Pfau}}]{Pfau_2019}%
		\BibitemOpen
		\bibfield  {author} {\bibinfo {author} {\bibfnamefont {M.}~\bibnamefont
				{Guo}}, \bibinfo {author} {\bibfnamefont {F.}~\bibnamefont {B{\"o}ttcher}},
			\bibinfo {author} {\bibfnamefont {J.}~\bibnamefont {Hertkorn}}, \bibinfo
			{author} {\bibfnamefont {J.-N.}\ \bibnamefont {Schmidt}}, \bibinfo {author}
			{\bibfnamefont {M.}~\bibnamefont {Wenzel}}, \bibinfo {author} {\bibfnamefont
				{H.~P.}\ \bibnamefont {B{\"u}chler}}, \bibinfo {author} {\bibfnamefont
				{T.}~\bibnamefont {Langen}}, \ and\ \bibinfo {author} {\bibfnamefont
				{T.}~\bibnamefont {Pfau}},\ }\href {\doibase 10.1038/s41586-019-1569-5}
		{\bibfield  {journal} {\bibinfo  {journal} {Nature}\ }\textbf {\bibinfo
				{volume} {574}},\ \bibinfo {pages} {386} (\bibinfo {year}
			{2019})}\BibitemShut {NoStop}%
		\bibitem [{\citenamefont {Tanzi}\ \emph {et~al.}(2019)\citenamefont {Tanzi},
			\citenamefont {Lucioni}, \citenamefont {Fam\`a}, \citenamefont {Catani},
			\citenamefont {Fioretti}, \citenamefont {Gabbanini}, \citenamefont {Bisset},
			\citenamefont {Santos},\ and\ \citenamefont {Modugno}}]{Modugno_2019}%
		\BibitemOpen
		\bibfield  {author} {\bibinfo {author} {\bibfnamefont {L.}~\bibnamefont
				{Tanzi}}, \bibinfo {author} {\bibfnamefont {E.}~\bibnamefont {Lucioni}},
			\bibinfo {author} {\bibfnamefont {F.}~\bibnamefont {Fam\`a}}, \bibinfo
			{author} {\bibfnamefont {J.}~\bibnamefont {Catani}}, \bibinfo {author}
			{\bibfnamefont {A.}~\bibnamefont {Fioretti}}, \bibinfo {author}
			{\bibfnamefont {C.}~\bibnamefont {Gabbanini}}, \bibinfo {author}
			{\bibfnamefont {R.~N.}\ \bibnamefont {Bisset}}, \bibinfo {author}
			{\bibfnamefont {L.}~\bibnamefont {Santos}}, \ and\ \bibinfo {author}
			{\bibfnamefont {G.}~\bibnamefont {Modugno}},\ }\href {\doibase
			10.1103/PhysRevLett.122.130405} {\bibfield  {journal} {\bibinfo  {journal}
				{Phys. Rev. Lett.}\ }\textbf {\bibinfo {volume} {122}},\ \bibinfo {pages}
			{130405} (\bibinfo {year} {2019})}\BibitemShut {NoStop}%
		\bibitem [{\citenamefont {W\"achtler}\ and\ \citenamefont
			{Santos}(2016)}]{Santos_2016}%
		\BibitemOpen
		\bibfield  {author} {\bibinfo {author} {\bibfnamefont {F.}~\bibnamefont
				{W\"achtler}}\ and\ \bibinfo {author} {\bibfnamefont {L.}~\bibnamefont
				{Santos}},\ }\href {\doibase 10.1103/PhysRevA.94.043618} {\bibfield
			{journal} {\bibinfo  {journal} {Phys. Rev. A}\ }\textbf {\bibinfo {volume}
				{94}},\ \bibinfo {pages} {043618} (\bibinfo {year} {2016})}\BibitemShut
		{NoStop}%
		\bibitem [{\citenamefont {Bisset}\ \emph {et~al.}(2016)\citenamefont {Bisset},
			\citenamefont {Wilson}, \citenamefont {Baillie},\ and\ \citenamefont
			{Blakie}}]{Blakie_2016}%
		\BibitemOpen
		\bibfield  {author} {\bibinfo {author} {\bibfnamefont {R.~N.}\ \bibnamefont
				{Bisset}}, \bibinfo {author} {\bibfnamefont {R.~M.}\ \bibnamefont {Wilson}},
			\bibinfo {author} {\bibfnamefont {D.}~\bibnamefont {Baillie}}, \ and\
			\bibinfo {author} {\bibfnamefont {P.~B.}\ \bibnamefont {Blakie}},\ }\href
		{\doibase 10.1103/PhysRevA.94.033619} {\bibfield  {journal} {\bibinfo
				{journal} {Phys. Rev. A}\ }\textbf {\bibinfo {volume} {94}},\ \bibinfo
			{pages} {033619} (\bibinfo {year} {2016})}\BibitemShut {NoStop}%
		\bibitem [{\citenamefont {Chomaz}\ \emph {et~al.}(2018)\citenamefont {Chomaz},
			\citenamefont {van Bijnen}, \citenamefont {Petter}, \citenamefont {Faraoni},
			\citenamefont {Baier}, \citenamefont {Becher}, \citenamefont {Mark},
			\citenamefont {W{\"a}chtler}, \citenamefont {Santos},\ and\ \citenamefont
			{Ferlaino}}]{Ferlaino_2018}%
		\BibitemOpen
		\bibfield  {author} {\bibinfo {author} {\bibfnamefont {L.}~\bibnamefont
				{Chomaz}}, \bibinfo {author} {\bibfnamefont {R.~M.~W.}\ \bibnamefont {van
					Bijnen}}, \bibinfo {author} {\bibfnamefont {D.}~\bibnamefont {Petter}},
			\bibinfo {author} {\bibfnamefont {G.}~\bibnamefont {Faraoni}}, \bibinfo
			{author} {\bibfnamefont {S.}~\bibnamefont {Baier}}, \bibinfo {author}
			{\bibfnamefont {J.~H.}\ \bibnamefont {Becher}}, \bibinfo {author}
			{\bibfnamefont {M.~J.}\ \bibnamefont {Mark}}, \bibinfo {author}
			{\bibfnamefont {F.}~\bibnamefont {W{\"a}chtler}}, \bibinfo {author}
			{\bibfnamefont {L.}~\bibnamefont {Santos}}, \ and\ \bibinfo {author}
			{\bibfnamefont {F.}~\bibnamefont {Ferlaino}},\ }\href {\doibase
			10.1038/s41567-018-0054-7} {\bibfield  {journal} {\bibinfo  {journal} {Nature
					Physics}\ }\textbf {\bibinfo {volume} {14}},\ \bibinfo {pages} {442}
			(\bibinfo {year} {2018})}\BibitemShut {NoStop}%
		\bibitem [{\citenamefont {Natale}\ \emph {et~al.}(2019)\citenamefont {Natale},
			\citenamefont {van Bijnen}, \citenamefont {Patscheider}, \citenamefont
			{Petter}, \citenamefont {Mark}, \citenamefont {Chomaz},\ and\ \citenamefont
			{Ferlaino}}]{Natale_2019}%
		\BibitemOpen
		\bibfield  {author} {\bibinfo {author} {\bibfnamefont {G.}~\bibnamefont
				{Natale}}, \bibinfo {author} {\bibfnamefont {R.~M.~W.}\ \bibnamefont {van
					Bijnen}}, \bibinfo {author} {\bibfnamefont {A.}~\bibnamefont {Patscheider}},
			\bibinfo {author} {\bibfnamefont {D.}~\bibnamefont {Petter}}, \bibinfo
			{author} {\bibfnamefont {M.~J.}\ \bibnamefont {Mark}}, \bibinfo {author}
			{\bibfnamefont {L.}~\bibnamefont {Chomaz}}, \ and\ \bibinfo {author}
			{\bibfnamefont {F.}~\bibnamefont {Ferlaino}},\ }\href {\doibase
			10.1103/PhysRevLett.123.050402} {\bibfield  {journal} {\bibinfo  {journal}
				{Phys. Rev. Lett.}\ }\textbf {\bibinfo {volume} {123}},\ \bibinfo {pages}
			{050402} (\bibinfo {year} {2019})}\BibitemShut {NoStop}%
		\bibitem [{\citenamefont {Landau}(1941)}]{landau_theory_1941}%
		\BibitemOpen
		\bibfield  {author} {\bibinfo {author} {\bibfnamefont {L.}~\bibnamefont
				{Landau}},\ }\href {\doibase 10.1103/PhysRev.60.356} {\bibfield  {journal}
			{\bibinfo  {journal} {Phys. Rev.}\ }\textbf {\bibinfo {volume} {60}},\
			\bibinfo {pages} {356} (\bibinfo {year} {1941})}\BibitemShut {NoStop}%
		\bibitem [{\citenamefont {Blakie}\ \emph
			{et~al.}(2020{\natexlab{a}})\citenamefont {Blakie}, \citenamefont {Baillie},\
			and\ \citenamefont {Pal}}]{Blakie_2020}%
		\BibitemOpen
		\bibfield  {author} {\bibinfo {author} {\bibfnamefont {P.~B.}\ \bibnamefont
				{Blakie}}, \bibinfo {author} {\bibfnamefont {D.}~\bibnamefont {Baillie}}, \
			and\ \bibinfo {author} {\bibfnamefont {S.}~\bibnamefont {Pal}},\ }\href
		{\doibase 10.1088/1572-9494/ab95fa} {\bibfield  {journal} {\bibinfo
				{journal} {Communications in Theoretical Physics}\ }\textbf {\bibinfo
				{volume} {72}},\ \bibinfo {pages} {085501} (\bibinfo {year}
			{2020}{\natexlab{a}})}\BibitemShut {NoStop}%
		\bibitem [{\citenamefont {Blakie}\ \emph
			{et~al.}(2020{\natexlab{b}})\citenamefont {Blakie}, \citenamefont {Baillie},
			\citenamefont {Chomaz},\ and\ \citenamefont {Ferlaino}}]{Ferlaino_2020}%
		\BibitemOpen
		\bibfield  {author} {\bibinfo {author} {\bibfnamefont {P.~B.}\ \bibnamefont
				{Blakie}}, \bibinfo {author} {\bibfnamefont {D.}~\bibnamefont {Baillie}},
			\bibinfo {author} {\bibfnamefont {L.}~\bibnamefont {Chomaz}}, \ and\ \bibinfo
			{author} {\bibfnamefont {F.}~\bibnamefont {Ferlaino}},\ }\href {\doibase
			10.1103/PhysRevResearch.2.043318} {\bibfield  {journal} {\bibinfo  {journal}
				{Phys. Rev. Research}\ }\textbf {\bibinfo {volume} {2}},\ \bibinfo {pages}
			{043318} (\bibinfo {year} {2020}{\natexlab{b}})}\BibitemShut {NoStop}%
		\bibitem [{Note2()}]{Note2}%
		\BibitemOpen
		\bibinfo {note} {Note that in~\protect \text {Fig.}~\ref {fig1} and~\protect
			\text {Fig.}~\ref {fig2-3} $g_d/g$ is varied by modifying the dipole moment
			as $7<d/\mu _B<10$ with $a$ constant. Though not
			realistic from an experimental point of view, this helps in a direct comparison of the interplay purely between the two long-range interactions.
			Nonetheless, it is conceivable that every point on the phase diagram can also be attained by varying $a$ for fixed values of $d$}\BibitemShut {NoStop}%
				\bibitem [{Note2a()}]{Note2a}%
		\BibitemOpen
		\bibinfo {note} {This \emph{bi}-roton instability owes its existence to the distinct functional form of the two interactions considered here. It is not an immediate consequence of having two long-range interactions - see Ref.~\cite{Ghosh2022} for a counter example}\BibitemShut {NoStop}%
		\bibitem{Ghosh2022}{R.~Ghosh, C.~Mishra, L.~Santos, and R.~Nath, \href{https://arxiv.org/abs/2210.01093}{arXiv:2210.01093 (2022)}.}

		\bibitem [{Note3()}]{Note3}%
		\BibitemOpen
		\bibinfo {note} {Note that within our current model, which neglects
			propagation effects in the Helmholtz equation, the usual thermodynamic limit
			$N,L \rightarrow \infty$ and finite $N/L$ leads to
			diverging energy~\cite {Ritsch_2016}. While the modified thermodynamic limit $L \rightarrow \infty $ with finite $N$~\cite {Ritsch_2016} nullifies the effect of dipole and contact
			interactions but keeps the light-induced interactions finite}\BibitemShut {NoStop}%
		\bibitem [{\citenamefont {Antoine}\ \emph {et~al.}(2017)\citenamefont
			{Antoine}, \citenamefont {Levitt},\ and\ \citenamefont {Tang}}]{Tang_2017}%
		\BibitemOpen
		\bibfield  {author} {\bibinfo {author} {\bibfnamefont {X.}~\bibnamefont
				{Antoine}}, \bibinfo {author} {\bibfnamefont {A.}~\bibnamefont {Levitt}}, \
			and\ \bibinfo {author} {\bibfnamefont {Q.}~\bibnamefont {Tang}},\ }\href
		{\doibase https://doi.org/10.1016/j.jcp.2017.04.040} {\bibfield  {journal}
			{\bibinfo  {journal} {Journal of Computational Physics}\ }\textbf {\bibinfo
				{volume} {343}},\ \bibinfo {pages} {92} (\bibinfo {year} {2017})}\BibitemShut
		{NoStop}%
		\bibitem [{\citenamefont {Ronen}\ \emph {et~al.}(2006)\citenamefont {Ronen},
			\citenamefont {Bortolotti},\ and\ \citenamefont {Bohn}}]{Bohn_2006}%
		\BibitemOpen
		\bibfield  {author} {\bibinfo {author} {\bibfnamefont {S.}~\bibnamefont
				{Ronen}}, \bibinfo {author} {\bibfnamefont {D.~C.~E.}\ \bibnamefont
				{Bortolotti}}, \ and\ \bibinfo {author} {\bibfnamefont {J.~L.}\ \bibnamefont
				{Bohn}},\ }\href {\doibase 10.1103/PhysRevA.74.013623} {\bibfield  {journal}
			{\bibinfo  {journal} {Phys. Rev. A}\ }\textbf {\bibinfo {volume} {74}},\
			\bibinfo {pages} {013623} (\bibinfo {year} {2006})}\BibitemShut {NoStop}%
		\bibitem [{\citenamefont {{Janner, A.}}\ \emph {et~al.}(1982)\citenamefont
			{{Janner, A.}}, \citenamefont {{Janssen, T.}},\ and\ \citenamefont {{de
					Wolff, P.M.}}}]{Wolff_1982}%
		\BibitemOpen
		\bibfield  {author} {\bibinfo {author} {\bibnamefont {{Janner, A.}}},
			\bibinfo {author} {\bibnamefont {{Janssen, T.}}}, \ and\ \bibinfo {author}
			{\bibnamefont {{de Wolff, P.M.}}},\ }\href {\doibase 10.1051/epn/19821312001}
		{\bibfield  {journal} {\bibinfo  {journal} {Europhys. News}\ }\textbf
			{\bibinfo {volume} {13}},\ \bibinfo {pages} {1} (\bibinfo {year}
			{1982})}\BibitemShut {NoStop}%
		\bibitem [{\citenamefont {Baillie}\ and\ \citenamefont
			{Blakie}(2018)}]{Blakie_2018}%
		\BibitemOpen
		\bibfield  {author} {\bibinfo {author} {\bibfnamefont {D.}~\bibnamefont
				{Baillie}}\ and\ \bibinfo {author} {\bibfnamefont {P.~B.}\ \bibnamefont
				{Blakie}},\ }\href {\doibase 10.1103/PhysRevLett.121.195301} {\bibfield
			{journal} {\bibinfo  {journal} {Phys. Rev. Lett.}\ }\textbf {\bibinfo
				{volume} {121}},\ \bibinfo {pages} {195301} (\bibinfo {year}
			{2018})}\BibitemShut {NoStop}%
		\bibitem [{\citenamefont {Leggett}(1970)}]{Leggett_1970}%
		\BibitemOpen
		\bibfield  {author} {\bibinfo {author} {\bibfnamefont {A.~J.}\ \bibnamefont
				{Leggett}},\ }\href {\doibase 10.1103/PhysRevLett.25.1543} {\bibfield
			{journal} {\bibinfo  {journal} {Phys. Rev. Lett.}\ }\textbf {\bibinfo
				{volume} {25}},\ \bibinfo {pages} {1543} (\bibinfo {year}
			{1970})}\BibitemShut {NoStop}%
		\bibitem [{\citenamefont {Sep\'ulveda}\ \emph {et~al.}(2008)\citenamefont
			{Sep\'ulveda}, \citenamefont {Josserand},\ and\ \citenamefont
			{Rica}}]{Rica_2008}%
		\BibitemOpen
		\bibfield  {author} {\bibinfo {author} {\bibfnamefont {N.}~\bibnamefont
				{Sep\'ulveda}}, \bibinfo {author} {\bibfnamefont {C.}~\bibnamefont
				{Josserand}}, \ and\ \bibinfo {author} {\bibfnamefont {S.}~\bibnamefont
				{Rica}},\ }\href {\doibase 10.1103/PhysRevB.77.054513} {\bibfield  {journal}
			{\bibinfo  {journal} {Phys. Rev. B}\ }\textbf {\bibinfo {volume} {77}},\
			\bibinfo {pages} {054513} (\bibinfo {year} {2008})}\BibitemShut {NoStop}%
			\bibitem [{\citenamefont {Bland}\ \emph {et~al.}(2022)\citenamefont {Bland},
  \citenamefont {Poli}, \citenamefont {Politi}, \citenamefont {Klaus},
  \citenamefont {Norcia}, \citenamefont {Ferlaino}, \citenamefont {Santos},\
  and\ \citenamefont {Bisset}}]{Bland2022}%
  \BibitemOpen
  \bibfield  {author} {\bibinfo {author} {\bibfnamefont {T.}~\bibnamefont
  {Bland}}, \bibinfo {author} {\bibfnamefont {E.}~\bibnamefont {Poli}},
  \bibinfo {author} {\bibfnamefont {C.}~\bibnamefont {Politi}}, \bibinfo
  {author} {\bibfnamefont {L.}~\bibnamefont {Klaus}}, \bibinfo {author}
  {\bibfnamefont {M.~A.}\ \bibnamefont {Norcia}}, \bibinfo {author}
  {\bibfnamefont {F.}~\bibnamefont {Ferlaino}}, \bibinfo {author}
  {\bibfnamefont {L.}~\bibnamefont {Santos}}, \ and\ \bibinfo {author}
  {\bibfnamefont {R.~N.}\ \bibnamefont {Bisset}},\ }\href {\doibase
  10.1103/PhysRevLett.128.195302} {\bibfield  {journal} {\bibinfo  {journal}
  {Phys. Rev. Lett.}\ }\textbf {\bibinfo {volume} {128}},\ \bibinfo
  {pages} {195302} (\bibinfo {year} {2022})}\BibitemShut {NoStop}%
		\bibitem [{\citenamefont {Ostermann}\ \emph {et~al.}(2022)\citenamefont
			{Ostermann}, \citenamefont {Walther},\ and\ \citenamefont
			{Yelin}}]{Yelin_2021}%
		\BibitemOpen
		\bibfield  {author} {\bibinfo {author} {\bibfnamefont {S.}~\bibnamefont
				{Ostermann}}, \bibinfo {author} {\bibfnamefont {V.}~\bibnamefont {Walther}},
			\ and\ \bibinfo {author} {\bibfnamefont {S.~F.}\ \bibnamefont {Yelin}},\
		}\href {\doibase 10.1103/PhysRevResearch.4.023074} {\bibfield  {journal}
			{\bibinfo  {journal} {Phys. Rev. Research}\ }\textbf {\bibinfo {volume}
				{4}},\ \bibinfo {pages} {023074} (\bibinfo {year} {2022})}\BibitemShut
		{NoStop}%
		\bibitem [{\citenamefont {Roccuzzo}\ \emph {et~al.}(2022)\citenamefont
			{Roccuzzo}, \citenamefont {Stringari},\ and\ \citenamefont
			{Recati}}]{Recati_2022}%
		\BibitemOpen
		\bibfield  {author} {\bibinfo {author} {\bibfnamefont {S.~M.}\ \bibnamefont
				{Roccuzzo}}, \bibinfo {author} {\bibfnamefont {S.}~\bibnamefont {Stringari}},
			\ and\ \bibinfo {author} {\bibfnamefont {A.}~\bibnamefont {Recati}},\ }\href
		{\doibase 10.1103/PhysRevResearch.4.013086} {\bibfield  {journal} {\bibinfo
				{journal} {Phys. Rev. Research}\ }\textbf {\bibinfo {volume} {4}},\ \bibinfo
			{pages} {013086} (\bibinfo {year} {2022})}\BibitemShut {NoStop}%
		\bibitem [{\citenamefont {Guo}\ \emph {et~al.}(2021)\citenamefont {Guo},
			\citenamefont {Kroeze}, \citenamefont {Marsh}, \citenamefont
			{Gopalakrishnan}, \citenamefont {Keeling},\ and\ \citenamefont
			{Lev}}]{Guo_2021}%
		\BibitemOpen
		\bibfield  {author} {\bibinfo {author} {\bibfnamefont {Y.}~\bibnamefont
				{Guo}}, \bibinfo {author} {\bibfnamefont {R.~M.}\ \bibnamefont {Kroeze}},
			\bibinfo {author} {\bibfnamefont {B.~P.}\ \bibnamefont {Marsh}}, \bibinfo
			{author} {\bibfnamefont {S.}~\bibnamefont {Gopalakrishnan}}, \bibinfo
			{author} {\bibfnamefont {J.}~\bibnamefont {Keeling}}, \ and\ \bibinfo
			{author} {\bibfnamefont {B.~L.}\ \bibnamefont {Lev}},\ }\href {\doibase
			10.1038/s41586-021-03945-x} {\bibfield  {journal} {\bibinfo  {journal}
				{Nature}\ }\textbf {\bibinfo {volume} {599}},\ \bibinfo {pages} {211}
			(\bibinfo {year} {2021})}\BibitemShut {NoStop}%
		\bibitem [{\citenamefont {Mishra}\ \emph {et~al.}(tion)\citenamefont {Mishra},
			\citenamefont {Ostermann}, \citenamefont {Mivehvar},\ and\ \citenamefont
			{Prasanna~Venkatesh}}]{FollowupPRA}%
		\BibitemOpen
		\bibfield  {author} {\bibinfo {author} {\bibfnamefont {C.}~\bibnamefont
				{Mishra}}, \bibinfo {author} {\bibfnamefont {S.}~\bibnamefont {Ostermann}},
			\bibinfo {author} {\bibfnamefont {F.}~\bibnamefont {Mivehvar}}, \ and\
			\bibinfo {author} {\bibfnamefont {B.}~\bibnamefont {Prasanna~Venkatesh}},\
		}\href@noop {} {\bibfield  {journal} {\bibinfo  {journal} {{}}\ } (\bibinfo
			{year} {In Preparation})}\BibitemShut {NoStop}%
		\bibitem [{\citenamefont {Steck}()}]{steck}%
		\BibitemOpen
		\bibfield  {author} {\bibinfo {author} {\bibfnamefont {D.~A.}\ \bibnamefont
				{Steck}},\ }\href@noop {} {\emph {\bibinfo {title} {Quantum and Atom
					Optics}}}\ (\bibinfo  {publisher} {available online at
			http://steck.us/teaching (revision 0.12.2, 11 April 2018)},\ \bibinfo {year}
		{{}})\BibitemShut {NoStop}%
		\bibitem [{\citenamefont {Lu}\ \emph {et~al.}(2011)\citenamefont {Lu},
			\citenamefont {Youn},\ and\ \citenamefont {Lev}}]{Lev_2011}%
		\BibitemOpen
		\bibfield  {author} {\bibinfo {author} {\bibfnamefont {M.}~\bibnamefont
				{Lu}}, \bibinfo {author} {\bibfnamefont {S.~H.}\ \bibnamefont {Youn}}, \ and\
			\bibinfo {author} {\bibfnamefont {B.~L.}\ \bibnamefont {Lev}},\ }\href
		{\doibase 10.1103/PhysRevA.83.012510} {\bibfield  {journal} {\bibinfo
				{journal} {Phys. Rev. A}\ }\textbf {\bibinfo {volume} {83}},\ \bibinfo
			{pages} {012510} (\bibinfo {year} {2011})}\BibitemShut {NoStop}%
	\end{thebibliography}
\end{document}